\documentclass[journal,comsoc]{IEEEtran}

\usepackage[T1]{fontenc}
\usepackage[mathcal]{euscript}
\usepackage{cite}
\usepackage[pdftex]{graphicx}
\usepackage{amsmath}
\usepackage{txfonts}
\usepackage{bbm}
\usepackage{algorithmicx}
\usepackage{algpseudocode}
\usepackage{algorithm}
\usepackage{array}
\usepackage[caption=false,font=footnotesize]{subfig}
\usepackage{pgfplots}
\usepackage{fixltx2e}
\pgfplotsset{compat=newest}
\usepackage{tikz}
\usetikzlibrary{positioning,shapes,shapes.geometric,calc,chains,shapes.multipart,arrows}
\tikzset{
	rubberduck/.style={
		shape=isosceles triangle,
		fill=red!30,
		minimum height=18pt,
		minimum width=12pt,
		shape border rotate=#1,
		isosceles triangle stretches,
		inner sep=0pt,
	},
	ap/.style={rubberduck=+90},
	queue/.style={
		rectangle split,
		minimum width=2em,
		rectangle split parts=4,
		draw,
		anchor=south,
	}
}

\usepackage{verbatim}

\makeatletter
\newenvironment{varsubequations}[1]
{%
  \addtocounter{equation}{-1}%
  \begin{subequations}%
  \renewcommand{\theparentequation}{#1}%
  \def\@currentlabel{#1}%
}{%
  \end{subequations}%
}
\makeatother

\newtheorem{theorem}{Theorem}

\newtheorem{proposition}{Proposition}

\DeclareMathOperator*{\minimize}{minimize}
\DeclareMathOperator*{\maximize}{maximize}
\DeclareMathOperator{\subjectto}{subject~to}

\algnewcommand\algorithmicinput{\textbf{INPUT:}}
\algnewcommand\INPUT{\item[\algorithmicinput]}

\algnewcommand\algorithmicoutput{\textbf{OUTPUT:}}
\algnewcommand\OUTPUT{\item[\algorithmicoutput]}

\algnewcommand\algorithmicinit{\textbf{Initialization:}}
\algnewcommand\Init{\item[\algorithmicinit]}

\algnewcommand\algorithmicpost{\textbf{Post Processing:}}
\algnewcommand\Post{\item[\algorithmicpost]}

\newcommand{\bs}{\boldsymbol{s}}
\newcommand{\br}{\boldsymbol{r}}
\newcommand{\bv}{\boldsymbol{v}}
\newcommand{\bw}{\boldsymbol{w}}
\newcommand{\bx}{\boldsymbol{x}}
\newcommand{\by}{\boldsymbol{y}}
\newcommand{\bz}{\boldsymbol{z}}
\newcommand{\bh}{\boldsymbol{h}}
\newcommand{\bF}{\boldsymbol{F}}
\newcommand{\bH}{\boldsymbol{H}}
\newcommand{\bG}{\boldsymbol{G}}
\newcommand{\balpha}{\boldsymbol{\alpha}}
\newcommand{\bbeta}{\boldsymbol{\beta}}
\newcommand{\bgamma}{\boldsymbol{\gamma}}
\newcommand{\Acal}{A}
\newcommand{\Kcal}{K}
\newcommand{\Lcal}{L}
\newcommand{\Ncal}{N}
\newcommand{\Ucal}{U}
\newcommand{\Links}{E}

\newcommand{\EW}[1]{#1}
\newcommand{\DG}[1]{#1}
\newcommand{\BZ}[1]{#1}
\newcommand{\BZr}[1]{{#1}}
\newcommand{\EWr}[1]{{#1}}
\newcommand{\BZf}[1]{{#1}}
\newcommand{\BZa}[1]{{#1}}

\newcommand{\ioj}[1]{_{#1}}
\newcommand{\aol}[1]{^{#1}}
\newcommand{\ijA}[4]{#1\ioj{#2\to#3}\aol{#4}}
\newcommand{\rj}[1]{r\ioj{#1}}
\newcommand{\sa}[3]{\ijA{s}{#1}{#2}{#3}}

\newcommand{\wi}[2]{w\ioj{#1}\aol{#2}}
\newcommand{\xa}[3]{\ijA{x}{#1}{#2}{#3}}
\newcommand{\yA}[2][]{y\ioj{#1}\aol{#2}}
\newcommand{\za}[3]{\ijA{z}{#1}{#2}{#3}}
\newcommand{\Aj}[1]{\Acal\ioj{#1}}
\newcommand{\Bl}[2][]{B\ioj{#1}\aol{#2}}
\newcommand{\Ni}[1]{\Ncal\ioj{#1}}
\newcommand{\Ui}[1]{\Ucal\ioj{#1}}

\begin{document}

\title{Large-Scale Spectrum Allocation for Cellular Networks via Sparse Optimization}

\author{Binnan~Zhuang,
	Dongning~Guo,~\IEEEmembership{Senior Member,~IEEE,}
	Ermin~Wei,
	and Michael~L.~Honig,~\IEEEmembership{Fellow,~IEEE}%
	\thanks{
	B.~Zhuang was with the Department of Electrical
  Engineering and Computer Science at Northwestern University,
  Evanston, IL, 60208.
  He is now with Samsung Semiconductor, Inc., Modem R\&D Lab, San Diego, CA.
	D.~Guo, E.~Wei and M.~L.~Honig are with the Department of Electrical
  Engineering and Computer Science at Northwestern University,
  Evanston, IL, 60208.
	}%
	\thanks{
	This work was supported in part by a gift from Futurewei Technologies and by the National Science Foundation under Grant No.~CCF-1423040.
	}
}

\maketitle

\begin{abstract}
This paper studies \BZf{joint spectrum allocation and user association}
in large heterogeneous cellular networks. The objective is to
maximize \BZf{some network utility function based on
given traffic statistics}
collected over a slow timescale, conceived to be seconds to minutes.
A key challenge is scalability:
interference across cells creates dependencies across the entire network,
making the optimization problem computationally challenging as the size of the
network becomes large. A suboptimal solution is presented, which performs
well in networks consisting of one hundred access points (APs)
serving several hundred \BZf{user devices}. This is achieved by
optimizing over local overlapping neighborhoods,
defined by interference conditions, and by exploiting the sparsity of a globally optimal solution.
Specifically, with a total of $k$ user devices in the entire network, it suffices to divide the spectrum into $k$ segments, where each segment is mapped to a particular set, or {\em pattern}, of active APs within each local neighborhood.
The problem is then to find a mapping of segments to patterns,
and to optimize the widths of the segments.
A convex relaxation is proposed for this, which relies on a re-weighted
$\ell_1$ approximation of an $\ell_0$ constraint, and is used to enforce
the mapping of a unique pattern to each spectrum segment.
A distributed implementation based on alternating direction method of multipliers (ADMM)
is also proposed. Numerical comparisons with benchmark schemes show that the proposed
method achieves a substantial increase in achievable throughput \BZf{and/or reduction in the average packet delay}.
\end{abstract}


\section{Introduction}
\label{sec:Intro}
Heterogeneous cellular networks with densely deployed access points (APs) have been proposed
for Long Term Evolution Advanced (LTE-A)~\cite{3GPP_TR-36.814,hwang2013holistic,nakamura2013trends},
and are anticipated to be key components of 5G networks.
The deployment of such dense networks brings new challenges with interference management.
Mitigating inter-cell interference, in particular, requires coordinated radio resource
allocation across multiple cells. Methods that operate over fast timescales include
multi-cell joint scheduling~\cite{yu2013multicell,frank2010cooperative,wang2013joint}
and dynamic spectrum allocation methods
associated with orthogonal frequency-division multiple access (OFDMA)~\cite{
stolyar2008self,
chang2009multicell, 
ali2009dynamic,
madan2010cell,
liao2014base,
fang2015joint
},
in combination with power control and beamforming.

The assignment of mobiles to APs, or user association, can also take into account
the interference environment. Those methods include assigning the user to the strongest AP
and the {\em range extension} techniques~\cite{KhaBhuJi10EW,DamMonYon11TransWC}
for balancing the load between macro and pico tiers,
along with more sophisticated optimizations of
an overall
\BZf{utility objective}~\cite{shen2014distributed,
fooladivanda2013joint,HonLuo13TransSP,kuang2012joint,LinYu13GLOBE,YeRonChe13TransWC}.

This paper considers the joint optimization of resource allocation and user association
in a large network with many APs. The objective is to optimize a network utility,
such as average delay, given traffic statistics and \BZr{average channel state information}
that change slowly over a geographic region. 
Our approach builds upon
the {\em slow}-timescale optimization
framework proposed in~\cite{
zhuang2015traffic-driven,
zhuang2016energy-efficient}.
``Slow'' refers to timescales over which average packet arrival and departure rates
are relatively stationary. \BZr{The timescale is conceived to be seconds to minutes in current networks.} Given a network of $n$ APs and $k$ mobile devices,
the spectrum is partitioned into $2^n$ {\em patterns},
corresponding to all possible subsets of active APs~\cite{zhuang2015traffic-driven}.
The problem, as originally formulated in~\cite{zhuang2015traffic-driven}, is then
to optimize the widths of spectrum segments, corresponding to the different patterns,
along with the association of patterns with devices.
The solution has been shown to provide significant performance improvement
in throughput enhancement, delay reduction, and energy savings\cite{
zhuang2015traffic-driven,zhuang2016energy-efficient,zhuang2017scalable,
kuang2016optimal,
kuang2016energy,zhou2017licensed,teng2015resource}.

Although it was shown in~\cite{zhuang2015traffic-driven} and~\cite{zhuang2016energy-efficient}
that the solution is {\em sparse} (at most $k$ out of $2^n$ patterns \BZr{have nonzero bandwidth}),
finding the set of optimal patterns that appear in the solution
is in general NP-hard. In prior work~\cite{zhuang2017scalable},
we have proposed a scalable approach to finding an approximate solution
by recognizing that each link rate depends only on a {\em local pattern},
containing only those APs within an {\em interference cluster}.
The problem can then be redefined over sets of overlapping \DG{clusters},
associated with those local pattern\DG{s}.
Each AP has its own interference \DG{cluster}, 
which captures the interference from nearby APs.
Additional constraints are needed to ensure that the
spectrum assigned to each particular AP is consistent
across \DG{all clusters}
to which it belongs. Even with those
constraints, however,
\DG{the convex optimization may not}
find consistent
{\em placements} of the spectrum \BZa{segments for all APs within} 
the available band. A discrete coloring algorithm is proposed
in~\cite{zhuang2017scalable} to ensure that the local patterns
are globally consistent. In this way,
the total number of variables is reduced from $O(2^n)$
in~\cite{zhuang2015traffic-driven} to polynomial in $n$,
facilitating scalability.

In this paper, we take a different approach to address the scalability \BZa{problem},
which exploits the fact that \BZr{there exists a globally optimal solution that}
contains at most $k$ \BZr{active} patterns.
Specifically, we reformulate the problem by dividing the spectrum
into $k$ segments, and attempt to identify
the pattern that should be associated with each segment.
This effectively reverses the approach in~\cite{zhuang2017scalable},
which attempts to assign a segment of spectrum to each pattern.
In this reformulation, we initially allow any combination of patterns
that can be assigned to each of the $k$ segments.
This problem is a convex relaxation of the original problem.
The one-to-one mapping of spectrum segments to patterns is then
enforced with an $\ell_0$ (cardinality) constraint.
An algorithm for finding an approximate solution to this problem
is presented based on a reweighted $\ell_1$ approximation
of the $\ell_0$ constraint~\cite{candes2008enhancing}.

The approach to scalability presented here has the following advantages
relative to the approach in~\cite{zhuang2017scalable}. First, it effectively
trades the combinatorial coloring problem that arises in~\cite{zhuang2017scalable}
with the $\ell_0$ constraint introduced here. Although this does not simplify
the original problem, it helps in finding an approximate solution,
since reweighted $\ell_1$ approximations for the $\ell_0$ norm
have been \BZa{known} to perform well. Second, the number of variables is reduced
from $O(2^n)$ in~\cite{zhuang2017scalable} to $O(nk)$, facilitating scalability.
Third, the numerical results presented here indicate that this method
generally gives better performance for a fixed computational complexity
than the method in~\cite{zhuang2017scalable}. Decomposing the centralized iterative algorithm into subproblems based on the alternating direction method of multipliers (ADMM)~\cite{boyd2011distributed}, we also develop a distributed solution.

In related work~\cite{shen2017fplinq}, the problem is to select
an active set of links (equivalently, a pattern)
on a particular time slot (equivalently, a frequency band)
in a peer-to-peer network
to maximize a weighted sum rate.
An iterative algorithm based on fractional programming
determines a {\em single} pattern for each time slot.
In contrast, we jointly optimize the patterns and their bandwidths.
Kuang et al~\cite{kuang2016optimal} considered a similar framework for optimizing spectrum allocation and user association, where the search set is limited to a small number of patterns {\em a priori} to avoid the combinatorial complexity.

The rest of the paper is organized as follows.
The system model is presented in Section~\ref{sec:SysMod}.
The original formulation with $2^n$ global patterns is presented
in Section~\ref{sec:GlobalForm}.
A scalable formulation with a sparsity constraint is presented
in Section~\ref{sec:MainRes}. An efficient centralized iterative algorithm
for finding an approximate solution is presented in Section~\ref{sec:l1} and
a distributed algorithm based on ADMM is presented in Section~\ref{sec:ADMM}.
Simulation results are presented in Section~\ref{sec:NumRes}.
Concluding remarks are given in Section~\ref{sec:Con}.

\section{System Model}
\label{sec:SysMod}

\begin{figure}
	\centering
  \begin{tikzpicture}[x=1cm,y=1cm,node distance=0 cm,outer sep=0pt]
    \tiny
    \tikzstyle{specb}=[rectangle,draw,anchor=south west,text centered,minimum height=7mm,fill=blue!50]
    \tikzstyle{specg}=[rectangle,draw,anchor=south west,text centered,minimum height=7mm,fill=green!50]
    \tikzstyle{spec}=[rectangle,draw,anchor=south west,text centered,minimum height=7mm,fill=gray!50]

    \draw[color=white] (0,0) -- (0,2.1) -- (7.6,2.1) -- (7.6,0) -- (0,0);

\draw (0,0) -- (0,2.1) -- (7.6,2.1) -- (7.6,0) -- (0,0);
\node at (-0.4,0.3) {\small AP 3};
\node at (-0.4,1) {\small AP 2};
\node at (-0.4,1.7) {\small AP 1};
\node at (4,-.75) {\small frequency/time resources};

\node[spec,minimum width=6mm] (1_1) at (0,1.4) {};

\draw[decorate, decoration={brace,mirror}, blue] (0.025,-.03) -- (.575,-.03);
\node at (0.3,-.4) {\small $y^{\{1\}}$};

\node[spec,minimum width=10mm] (2_2) [below right = of 1_1]{};

\draw[decorate, decoration={brace,mirror}, blue] (.625,-.03) -- (1.575,-.03);
\node at (1.1,-.4) {\small $y^{\{2\}}$};

\node[spec,minimum width=6mm] (3_3) [below right = of 2_2]{};

\draw[decorate, decoration={brace,mirror}, blue] (1.625,-.03) -- (2.175,-.03);
\node at (1.8,-.4) {\small $y^{\{3\}}$};

\node[spec,minimum width=9mm] (12_1) at (2.2,1.4) {};
\node[spec,minimum width=9mm] (12_2) [above right = of 3_3]{};

\draw[decorate, decoration={brace,mirror}, blue] (2.225,-.03) -- (3.075,-.03);
\node at (2.7,-.4) {\small $y^{\{1,2\}}$};

\node[spec,minimum width=11mm] (23_2) [right = of 12_2]{};
\node[spec,minimum width=11mm] (23_3) [below right = of 12_2]{};

\draw[decorate, decoration={brace,mirror}, blue] (3.125,-.03) -- (4.175,-.03);
\node at (3.7,-.4) {\small $y^{\{2,3\}}$};

\node[spec,minimum width=21mm] (13_1) [above right = of 23_2]{};
\node[spec,minimum width=21mm] (13_3) [below right = of 23_2]{};

\draw[decorate, decoration={brace,mirror}, blue] (4.225,-.03) -- (6.275,-.03);
\node at (5.2,-.4) {\small $y^{\{1,3\}}$};

\node[spec,minimum width=13mm] (123_1) [right = of 13_1]{};
\node[spec,minimum width=13mm] (123_2) [below right = of 13_1]{};
\node[spec,minimum width=13mm] (123_3) [right = of 13_3]{};

\draw[decorate, decoration={brace,mirror}, blue] (6.325,-.03) -- (7.575,-.03);
\node at (6.8,-.4) {\small $y^{\{1,2,3\}}$};

\node[specb,minimum width=3mm] (1_1-1) at (0,1.4) {1};
\node[specg,minimum width=3mm] (1_1-2) [right = of 1_1-1]{2};
\node[specb,minimum width=5mm] (2_2-1) [below right = of 1_1-2]{1};
\node[specg,minimum width=5mm] (2_2-2) [right = of 2_2-1]{2};
\node[specb,minimum width=3mm] (3_3-1) [below right = of 2_2-2]{1};
\node[specg,minimum width=3mm] (3_3-2) [right = of 3_3-1]{2};

\node[specb,minimum width=6mm] (12_1-1) at (2.2,1.4) {1}; 
\node[specg,minimum width=3mm] (12_1-2) [right = of 12_1-1]{2};
\node[specb,minimum width=3mm] (12_2-1) [above right = of 3_3-2]{1};
\node[specg,minimum width=6mm] (12_2-2) [right = of 12_2-1]{2};
\node[specb,minimum width=6mm] (23_2-1) [right = of 12_2-2]{1};
\node[specg,minimum width=5mm] (23_2-2) [right = of 23_2-1]{2};
\node[specb,minimum width=5mm] (23_3-1) [below right = of 12_2-2]{1};
\node[specg,minimum width=6mm] (23_3-2) [right = of 23_3-1]{2};
\node[specb,minimum width=9mm] (13_1-1) [above right = of 23_2-2]{1};
\node[specg,minimum width=12mm] (13_1-2) [right = of 13_1-1]{2};
\node[specb,minimum width=12mm] (13_3-1) [below right = of 23_2-2]{1};
\node[specg,minimum width=9mm] (13_3-2) [right = of 13_3-1]{2};

\node[specb,minimum width=5mm] (123_1-1) [right = of 13_1-2]{1};
\node[specg,minimum width=8mm] (123_1-2) [right = of 123_1-1]{2};
\node[specb,minimum width=9mm] (123_2-1) [below right = of 13_1-2]{1};
\node[specg,minimum width=4mm] (123_2-2) [right = of 123_2-1]{2};
\node[specb,minimum width=6mm] (123_3-1) [right = of 13_3-2]{1};
\node[specg,minimum width=7mm] (123_3-2) [right = of 123_3-1]{2};

\draw[decorate, decoration={brace}, blue] (4.225,2.12) -- (5.075,2.12);
\draw[decorate, decoration={brace}, blue] (5.125,2.12) -- (6.275,2.12);
\draw[decorate, decoration={brace}, blue] (4.225,0.72) -- (5.375,.72);
\draw[decorate, decoration={brace}, blue] (5.425,0.72) -- (6.275,.72);
\node at (4.6,2.5) {\small $\xa{1}{1}{\{1,3\}}$};
\node at (5.7,2.5) {\small $\xa{1}{2}{\{1,3\}}$};
\node at (4.8,1.05) {\small $\xa{3}{1}{\{1,3\}}$};
\node at (5.9,1.05) {\small $\xa{3}{2}{\{1,3\}}$};

\draw (.5,5) node[ap] (apone) {\small AP1};
\draw (3.5,5) node[ap] (aptwo) {\small AP2};
\draw (6.5,5) node[ap] (apthree) {\small AP3};
\node[draw,fill=blue!50] (ueone) at (2,3.5) {\small device1};
\node[draw,fill=green!50] (uetwo) at (5,3.5) {\small device2};
\draw [->] (apone) -- (ueone);
\draw [->] (aptwo) -- (ueone);
\draw [->] (aptwo) -- (uetwo);
\draw [->] (apthree) -- (uetwo);
\draw [->] (apthree) -- (ueone);
\draw [->] (apone) -- (uetwo);
\end{tikzpicture}
	\caption{Illustration of all patterns of a 3-AP 2-mobile network with spectrum allocation variables.}
	\label{fig:Mod}
\end{figure}
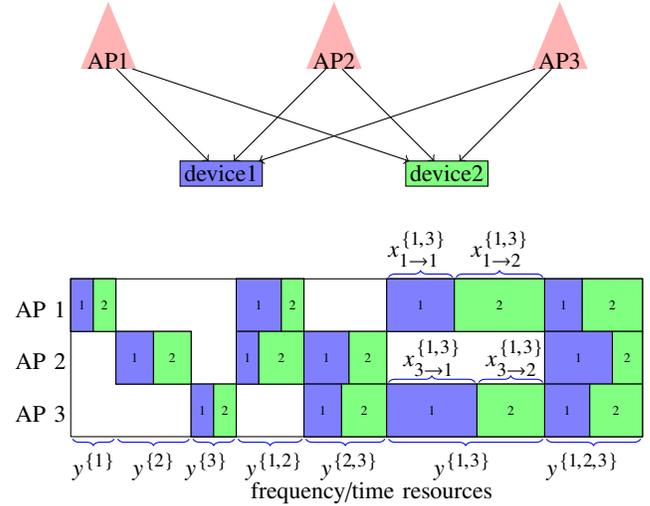

In a network with $n$ APs, we denote the set of AP indices as $\Ncal =\{1,\cdots,n\}$.  The $n$ APs share $W$ Hz of spectrum (treated as one unit), which can be considered homogeneous on a slow timescale.
(All segments of the resource have the same quality.)
\BZr{The notion of resource can also be generalized to time (scheduling) and \BZa{the combination of} spectrum and time (resource block allocation in the time-frequency grid). We focus on one time period on the slow timescale.}
Each AP can transmit on any part(s) of the spectrum. \EW{Hence, each slice of the spectrum can be shared by any subset of APs. We \DG{refer to} 
the $2^n$ possible ways (subsets of $n$ APs) to share a slice of the spectrum as 
{\it patterns}, each \DG{corresponding to a subset} $A\subset\Ncal$.}  If a slice of spectrum is designated to pattern $A$, only APs in $A$ can transmit over the spectrum \EW{and 
they interfere at \DG{the} devices they serve.%
}
\DG{Due to spectrum homogeneity, it}
suffices to describe
\DG{an allocation}
of the spectrum
\DG{to APs}
as $\{\yA{A}\}_{A\subset \Ncal }$, where $\yA{A}$ denotes the fraction of total bandwidth allocated to pattern $A$.
The total bandwidth allocated to all patterns (including the empty set $\emptyset$)
is one unit:
\begin{align}\label{eq:bandwidth}
  \sum_{A\subset \Ncal }\yA{A}=1.
\end{align}
An example with three APs
\DG{and two devices}
is depicted in Fig.~\ref{fig:Mod}. AP 1 exclusively owns \EW{the spectrum allocated to pattern} ${\{1\}}$; shares \BZ{the spectrum allocated to pattern} ${\{1,2\}}$ with AP 2; shares \BZ{the spectrum allocated to pattern} ${\{1,3\}}$ with AP 3; and shares \BZ{the spectrum allocated to pattern} ${\{1,2,3\}}$ with both AP 2 and AP 3.

\DG{In principle, a device may be served by any subset of APs and an AP may serve any subset of devices.  In practice, however, a device is only served by APs within a small neighborhood around it.  This is because the channel gain from a transmitter to a receiver vanishes quickly with}
\EW{the distance between them}. 
\DG{Let $\Kcal=\{1,\dots,k\}$ denote the set of $k$ device indices.}
Let $\Links\subset \Ncal \times \Kcal$ denote the set of \DG{(admissible)} links \EW{from APs to devices}.
\DG{The APs, mobile devices, and} links in $\Links$ form a bipartite graph with APs on one side and devices on the other.
Device $j$ can only be served by \DG{the set of neighboring} APs
\EW{in this bipartite graph, \DG{denoted} 
by set} 
\begin{align} \label{eq:SetA}
  \Aj{j}=\{i\in\Ncal \,|\, (i\to j)\in E\},
\end{align}
\BZr{where $i\to j$ denotes the link from AP $i$ to device $j$.} Likewise, AP $i$ can only serve its set of neighboring devices,
denoted by 
\begin{align} \label{eq:SetU}
  \Ui{i}=\{j\in\Kcal \,|\, (i\to j)\in E\}.
\end{align}
Evidently, $i\in A_j$ if and only if $j\in U_i$. Since an isolated AP or device in the graph should not be \BZa{assigned any resource at an efficient spectrum allocation}, we assume
\DG{no terminal is isolated without loss of generality, i.e., the sets $\Aj{j}$ and $\Ui{i}$ are nonempty.}

When $i\in A\cap A_j$, we use $\xa{i}{j}{A}$ to denote {the fraction of total bandwidth} used by AP $i$ to serve device $j$ under pattern $A$.
(Several such $x$-variables are illustrated in Fig.~\ref{fig:Mod}.)  Note that we do not define the variable if $i\notin A$ (AP $i$ does not transmit under pattern $A$) or $i\notin A_j$ (AP $i$ cannot serve device $j$).  Although we could equivalently set the variable to zero whenever $i\notin A\cap A_j$, we simply
omit those variables.
\BZr{We assume that an AP uses orthogonal (non-overlapping) spectrum to serve different devices.}
Since the total bandwidth assigned to pattern $A$ is $\yA{A}$, we have:
\begin{align} \label{eq:sum_x=y}
  \sum_{j\in U_i }\xa{i}{j}{A}= \yA{A},~\forall A\subset\Ncal , i\in A.
\end{align}

\DG{For $i\in A\cap A_j$,}
let $\sa{i}{j}{A}$ denote the \DG{{\em value}} of the link from AP $i$ to device $j$
\DG{per unit of resource under pattern $A$.
Again, the parameter is undefined if $i\notin A\cap A_j$ (as opposed to setting it to zero).}
\DG{For concreteness, we let the coefficient $\sa{i}{j}{A}$ represent the spectral efficiency of link $i\to j$ under pattern $A$.}
We assume
\DG{that when AP $i$ transmits over any part(s) of the spectrum, it transmits with fixed flat power spectral density (PSD) $p_i$.\footnote{Power control will be considered in future work.}}
The parameter $\sa{i}{j}{A}$ is determined by the \BZ{pathloss}
\DG{and} shadowing \DG{of link $i\to j$}, and 
characteristics of the interference links from \DG{other} APs in $A$
to device $j$.
\DG{In this paper, we assume that when a device decodes information from one AP's signals over a slice of spectrum, it treats all interference over the same spectrum as noise.\footnote{The framework can be generalized to treat many forms of coordinated multipoint (CoMP) transmissions.  For example, cooperative transmission is considered in~\cite{li2017cloud-based}.}}
For concreteness, we use Shannon's formula to \DG{write: 
\begin{align} \label{eq:SE}
  \sa{i}{j}{A}
  =
  W\, \log_2\left(1+\frac{p_i g\ioj {i\to j}}{\sum_{i'\in A\setminus\{i\}}p_{i'}g\ioj {i'\to j}+n_j}\right), \; \forall i\in A\cap A_j
\end{align}
in bits per second}, where
$g\ioj {i\to j}$
\DG{denotes the (slow-timescale average)}
power gain of link
{$i\to j$}, and $n_j$ is the noise PSD at device $j$.
With this definition the total data rate
\DG{for} device $j$ is then given by
\begin{align} \label{eq:rate}
  \rj j
	=
	\sum_{A\subset \Ncal }\sum_{i\in A\cap A_j}
	\sa{i}{j}{A}\xa{i}{j}{A},~\forall j\in\Kcal .
\end{align}
\BZr{Both the spectral efficiency and the service rate represent averages over the slow timescale.}

\DG{Given a specific allocation}, if AP $i$ transmits to device $j$ under at least one pattern, i.e.,
\DG{$i\in A_j$ and} $\sum_{A\subset\Ncal:\,i\in A} \xa{i}{j}{A}>0$, then they are said to be {\it associated} with each other.

\section{Problem Formulation Using Global Patterns}
\label{sec:GlobalForm}

The objective of the slow-timescale optimization is to maximize a network utility \DG{function}
over the spectrum allocation
across all links \BZa{and patterns}
 \EW{represented} by
$\bx=\big(\xa{i}{j}{A}\big)_{A\subset \Ncal, i\in A, \DG{j\in U_i}}$. 
\DG{Let $u(r_1,\dots,r_k)$ denote a network utility function of the rate tuple $\br=[r_1,\dots,r_k]$.} \BZr{In each time period,} the optimization problem \EW{with constraints 
\DG{\eqref{eq:bandwidth}, \eqref{eq:sum_x=y}, and \eqref{eq:rate}}
is formulated} as~\ref{eq:Opt}:
\begin{varsubequations}{P0} \label{eq:Opt}
\begin{align}
  \maximize_{\br,\,\bx,\,\by}~
  & u(\rj 1,\cdots,\rj k) \label{eq:opt:obj}\\
  \subjectto~
  & \rj j=\sum_{A\subset\Ncal}\sum_{i\in A\cap A_j} \sa{i}{j}{A}\xa{i}{j}{A},
  \quad \forall j\in\Kcal \label{eq:opt:r=x}\\
  & \sum_{j\in U_i} \xa{i}{j}{A} = \yA{A},
  \quad \forall A\subset\Ncal,\;i\in A\label{eq:opt:x<=y}\\
  & \sum_{A\subset\Ncal} \yA{A} = 1 \label{eq:opt:y}\\
  & \xa{i}{j}{A}\geq 0,
  \quad \forall A\subset \Ncal,\; i\in A,\; j\in U_i . \label{eq:opt:x}
\end{align}
\end{varsubequations}%
\DG{Since all the constraints are linear,}
the optimization problem \DG{is} convex if $u(\br)$ is concave in
\DG{the service rate vector $\br$.}
The class of utility functions that make~\ref{eq:Opt} convex include \DG{the frequently-used (weighted)} sum rate, \DG{the} sum log-rate, and \DG{the} minimum user rate, among others.

\DG{For concreteness, we follow~\cite{zhuang2017scalable} and focus} on minimizing the average packet delay.
\DG{Specifically, we assume homogeneous Poisson packet arrivals for device $j$ at rate $\lambda_j$ and exponentially distributed packet lengths of $\tau$ bits on average.  The service rate $r_j/\tau$ (in packets/second) is sustainable for serving device $j$'s queue regardless of the state of other devices' queues.  In this case, device $j$'s queueing dynamics are precisely modeled by an M/M/1 queue. The average packet sojourn time is given by
\begin{align}
	\frac1{(\rj j/\tau-\lambda\ioj j)^+}
\end{align}%
}%
where \DG{$1/x^+=1/x$ if $x>0$ and $1/x^+=+\infty$ if $x\le0$}.\footnote{Unlike $1/x$, the function $1/x^+$ is convex on $(-\infty,\infty)$ and is recognized as such by many optimization software packages.
We therefore write $1/x^+$ instead of $1/x$, which is nonconvex on $(-\infty,\infty)$ with the constraint $x>0$.}
The network utility function \DG{can be expressed as:} 
\begin{align}\label{eq:AvgDelay}
  u(\rj 1,\cdots,\rj k)
	=
	-\sum_{j=1}^k\frac{\lambda\ioj j}{(\rj j/\tau-\lambda\ioj j)^+} .
\end{align}


\BZr{The traffic arrival rates and spectral efficiencies are updated once each period on the slow timescale. \ref{eq:Opt} is intended to be solved once each period on a slow timescale. The optimized patterns are used throughout the decision period of seconds or minutes. In fact, the notion of a device on such a slow timescale can be considered as \BZa{a} set of service requests from the same geographic area, which share the same quality of service (QoS) (due to the same long-term average spectral efficiencies). The pattern based spectrum allocation determines the spectrum needed to serve different types of service requests (from different locations). Thus, slow timescale spectrum allocation complements fast time resource allocation, i.e., implemented over a period measured in milliseconds. The interaction of the two timescales are discussed in~\cite{teng2015resource}.}

\DG{It is instructive to count the number of variables in~\ref{eq:Opt}.  It is not difficult to see that there are $k$ $r$-variables, $2^n$ $y$-variables, and the number of $x$-variables is
\begin{align}
	2^{n-1} \sum^n_{i=1} |U_i|
\end{align}
where $|\cdot|$ yields the cardinality of a set.  Even if the number of devices any AP can serve is upper bounded by a constant $k_0$ (i.e., $|U_i|\le k_0$, $\forall i\in N$), the total number of variables in~\ref{eq:Opt} is on the order of $O(n 2^n)$.  This suggests that, even though~\ref{eq:Opt} is convex, it is very hard to solve directly for all but a small number of APs.

To make progress, we shall use the following fact that~\ref{eq:Opt} admits a sparse optimal solution:}
\begin{proposition}\textrm{(\cite{zhuang2016energy-efficient})}   \label{thm:Spec}
  \DG{If the} utility function $u(\br)$ \DG{of~\ref{eq:Opt} is concave in the rate vector $\br$, then there exists a $(k+1)$-sparse optimal allocation, namely,} an optimal solution that
	\DG{satisfies:}
  \begin{align} \label{eq:ThmSpec}
    \left|\left\{ A\subset\Ncal~\big|~\yA{A}>0 \right\}\right|\leq k+1.
  \end{align}
  \DG{Furthermore, if} the utility function is element-wise nondecreasing in the rate vector, then there exists 
	\DG{a $k$-sparse}
	optimal solution.
\end{proposition}

\DG{By Proposition~\ref{thm:Spec}, it suffices to identify no more than $k+1$ out of the $2^n$ patterns to activate.}
This property
\DG{is the key to the} scalable algorithm
\DG{developed in the next section.}

\section{Reformulation with Sparsity Constraints}
\label{sec:MainRes}

Solving~\ref{eq:Opt} directly \DG{is} prohibitively expensive \DG{for large networks} 
due to the inherit complexity from the \DG{exponential number of} patterns,
\DG{referred to as {\em global patterns} in the sequel.  Although}
\EW{Proposition~\ref{thm:Spec} guarantees the existence of a sparse optimal solution, 
it remains} \DG{computationally difficult to determine a small subset} \EW{of active patterns at an optimal solution} out of the $2^n$ \DG{possible} patterns.
In this section, we
\DG{introduce the notion of}
{\em local pattern} and a 
relaxation \DG{to significantly reduce the number of variables.  We then exploit} the sparse structure of the optimal solution to derive a scalable formulation.

\subsection{Local Pattern}
\label{subsec:local}

\DG{The key idea is to approximate the link spectral efficiency under a global pattern by that under a local pattern, where the APs outside the local pattern (referred to as remote APs) are treated as stationary noise sources, \BZf{whose on/off dynamics can be ignored}.
For \BZ{device $j$}, its {\it local patterns} consist of all subsets of \BZf{its} neighborhood \BZ{$A_j$}.  Here
we assume all remote APs ($N\setminus A_j$) are always on, and generate interference.
This gives the pessimistic approximation
\begin{align} \label{eq:stilde}
  \ijA{\tilde{s}}{i}{j}{A} =
  \sa{i}{j}{A\cup(\Ncal\setminus A_j)}
  ,  \quad \forall i\in A\cap A_j.  
\end{align}
That is,
the spectral efficiency
is regarded to be identical to that under the global pattern \BZ{$A\cup(\Ncal\setminus A_j)$}, which includes all remote APs.
In contrast, an optimistic approximation is defined by ignoring all remote interference:
\BZ{$\ijA{\tilde{s}}{i}{j}{A}=\sa{i}{j}{A\cap A_j}$, $\forall i\in A\cap A_j$.}
There are, of course, other possibilities in between, e.g., reducing the amount of interference from remote APs according to their utilizations.
We note that if the neighborhoods are sufficiently large, so that remote APs' total interference is negligible compared to thermal noise, then the preceding approximations are arbitrarily accurate.
Indeed, if all links outside the set $E$ have zeros gains,
then the 
approximations become precise.
In this paper, we adopt the pessimistic assumption~\eqref{eq:stilde} \EW{for the} numerical results in Section~\ref{sec:NumRes}.}

\DG{
In the sequel, we abuse the notation slightly to redefine $\sa{i}{j}{A}$ in the case where $A$ does not include all remote APs:
\begin{align} \label{eq:ss}
 \sa{i}{j}{A} = \sa{i}{j}{A\cup(\Ncal\setminus \BZ{A_j})},
 \quad \text{if } i\in A\cap A_j \text{ and } N\setminus A_j \not\subset A .
\end{align}
Evidently,~\eqref{eq:ss} degrades those redefined efficiencies \BZa{in general}. Moreover, those redefined spectral efficiencies are equal to the corresponding local ones \BZf{defined in~\eqref{eq:stilde}}.
We henceforth use the notation $\sa{i}{j}{A}$ to represent the (redefined) spectral efficiencies under both global patterns ($A\subset\Ncal$) and local patterns (\BZ{$A\subset A_j$}).  A consequence of~\eqref{eq:ss} is that the spectral efficiency now depends only on the local pattern:}
\begin{align} \label{eq:slocal}
 \sa{i}{j}{A} = \sa{i}{j}{A\cap \BZ{A_j}},
 \quad \forall i\in A \cap A_j.  
\end{align}

\subsection{Local Allocation Variables}

\def\Uone{(-1.25,0.45) ellipse [x radius=1, y radius=0.4]}
\def\Uthree{(1.25,0.45) ellipse [x radius=1, y radius=0.4]}
\def\Utwo{(0,0.45) ellipse [x radius=2.5, y radius=0.55]}
\def\None{(-1.1,1.6) ellipse [x radius=1.65, y radius=0.45]}
\def\Nthree{(1.1,1.6) ellipse [x radius=1.65, y radius=0.45]}
\def\Ntwo{(0,1.8) ellipse [x radius=3.2, y radius=0.8]}

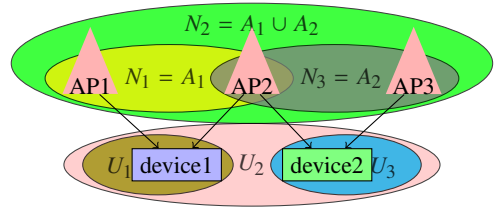
\begin{figure}
\centering
\small
\begin{tikzpicture}
 \filldraw[fill=green,opacity=0.75] \Ntwo node [label={[xshift=0, yshift=4.3]$\Ncal_2=\Acal_1\cup\Acal_2$}] {};
 \filldraw[fill=yellow,opacity=0.75] \None node (deltaone) {$\Ncal_1=\Acal_1\;$};
 \filldraw[fill=gray,opacity=0.75] \Nthree node {$\;\;\Ncal_3=\Acal_2$};
 \draw (-2.15,1.5) node[ap] (apone) {\small{AP1}};
 \draw (0,1.5) node[ap] (aptwo) {\small{AP2}};
 \draw (2.15,1.5) node[ap] (apthree) {\small{AP3}};
 \filldraw[fill=pink,opacity=0.75] \Utwo node [label={[xshift=0cm, yshift=0cm]}]{$\Ucal_2$};
 \filldraw[fill=olive,opacity=0.75] \Uone node (psione) [label={[xshift=-.5cm, yshift=-.4cm]$\Ucal_1$}] {};
 \filldraw[fill=cyan,opacity=0.75] \Uthree node (psithree)  [label={[xshift=.50cm, yshift=-.4cm]$\Ucal_3$}] {};
 \draw (-1,0.45) node[draw,fill=blue!30] (ueone) {\small{device1}};
 \draw (1,0.45) node[draw,fill=green!50] (uetwo) {\small{device2}};
 \draw [->] (apone) -- (ueone);
 \draw [->] (aptwo) -- (ueone);
 \draw [->] (aptwo) -- (uetwo);
 \draw [->] (apthree) -- (uetwo);
\end{tikzpicture}
\caption{Neighborhoods in the case of three APs and two devices.}
\label{fig:LocalPattern}
\end{figure}

We next introduce a set of local allocation variables, which shall be used to replace the original $O(n2^n)$ (global) variables.
\DG{Let us define}
the {\it interference cluster (or cluster)} of AP $i$ as:
\begin{align} \label{eq:SetN}
  \Ni{i} = \cup_{j\in\Ui{i}}\Aj{j},
\end{align}
which includes AP $i$ itself and all APs that may directly interfere with it. \DG{Fig.~\ref{fig:LocalPattern} depicts an example with three APs
and two devices.
The set of \DG{admissible} links 
are
$\Links=\{1\to 1,\;2\to 1,\;2\to 2,\;3\to 2\}$.
\EW{Therefore,} the AP neighborhoods are
$\Ucal_1=\{1\}$, $\Ucal_2=\{1,2\}$, and $\Ucal_3=\{2\}$;
and the device neighborhoods are $\Acal_1=\{1,2\}$ and $\Acal_2=\{2,3\}$. The interference cluster of AP 1 is $N_1=\{1,2\}$, as AP 2 interferes with it at device $1$.
The interference cluster of AP 3 is $N_3=\{2,3\}$, as AP 2 interferes with it at device $2$.
The interference cluster of AP 2 is $N_2=\{1,2,3\}$, since AP 1 and AP 3 interfere with it at device $1$ and device $2$,
respectively.}

In what follows, we assume \EW{that an AP serves at most $k_0$ devices and each device is served by at most $n_0$ APs, i.e.,} $|\Aj{j}|\le n_0$ for all $j\in\Kcal$ and $|\Ui{i}|\le k_0$ for all $i\in\Ncal$, where $k_0$ and $n_0$ are constants.  Thus the bipartite graph has finite node degrees. This implies an upper bound on the cluster sizes:
\begin{align}
  |\Ni{i}| \le k_0 n_0, \quad \forall i\in\Ncal.
\end{align}

\DG{We next rewrite the service rates defined in~\eqref{eq:rate} in terms of a new set of {\em local} allocation variables.
\BZ{Although the spectral efficiency of link $i\to j$ only depends on local patterns in the device neighborhood, \BZf{$A_j$,} the local allocation variables at AP $i$ are defined over all subsets of interference cluster $N_i$. This is because two local allocations in their respective interference clusters must be consistent over the overlapping area of the clusters, as shall be \BZf{illustrated} shortly.}
Specifically, for every admissible link $(i\to j)\in E$ and every subset $B$ of the cluster $N_i$ with $i\in B$, let
\begin{align} \label{eq:LocalGlobal}
  \za{i}{j}{B} = \sum_{A\subset\Ncal :A\cap\Ni{i}=B}\xa{i}{j}{A},
\end{align}
which represents the total bandwidth allocated to the link under all global patterns that match the local pattern $B$.
The total number of $z$-variables is:
\begin{align}
  \sum_{i\in\Ncal} |\Ui{i}| 2^{|\Ni{i}|-1}
  &\le n k_0 2^{k_0n_0} 
\end{align}
which grows linearly with the network size $n$.
For every $j\in\Kcal$, the service rate defined in~\eqref{eq:rate} and~\eqref{eq:opt:r=x} can be calculated as:
\begin{align}
  \rj j
	&= \sum_{i\in\Aj{j}}
		\sum_{A\subset N: i\in A}
		\sa{i}{j}{A} 
	 	\xa{i}{j}{A}\label{eq:riA}\\
	&= \sum_{i\in\Aj{j}}\,
		\sum_{B\subset\Ni{i}: i\in B} \sa{i}{j}{B} 
		\sum_{A\subset\Ncal: A\cap N_i=B}
		\xa{i}{j}{A}\label{eq:eq3}\\
  &= \sum_{i\in\Aj{j}} \sum_{B\subset N_i: i\in B} \sa{i}{j}{B} 
	 	\za{i}{j}{B}\label{eq:eq4}
\end{align}
where~\eqref{eq:eq3} follows from  
$\sa{i}{j}{A}
= \sa{i}{j}{A\cap A_j}
= \sa{i}{j}{A\cap N_i \cap A_j}
= \sa{i}{j}{B\cap A_j}
= \sa{i}{j}{B}$
whenever $A_j\subset N_i$ and $A \cap N_i = B$,\footnote{One may replace $\sa{i}{j}{B}$ by $\sa{i}{j}{B\cap A_j}$ in~\eqref{eq:eq4} to save storage needed for the spectral efficiency parameters.  We use $\sa{i}{j}{B}$ for ease of notation.}
and~\eqref{eq:eq4} is due to definition~\eqref{eq:LocalGlobal}.
The service rate $r_j$ now depends on no more than $O(n)$ $z$-variables in~\eqref{eq:eq4} in lieu of $O(n2^n)$ $x$-variables in~\eqref{eq:rate}.}

\subsection{An Equivalent Formulation}

\DG{In this subsection, we reformulate~\ref{eq:Opt} as an equivalent optimization problem using $O(kn)$ ``local'' variables in lieu of the $O(n2^n)$ global variables.  This is in part} motivated by the sparsity property \DG{guaranteed by} Proposition~\ref{thm:Spec}, i.e., \EW{there exists an optimal solution with at most $k+1$ active patterns}.  We reformulate the problem by dividing the spectrum into $k+1$ segments with the goal that 
\DG{each segment eventually corresponds to one pattern}
to activate.
\DG{The service rate to device $j$ is rewritten as:}
\begin{align} \label{eq:Con1}
  \rj j
  =
	\sum_{l\in L}
	\sum_{i\in\Aj{j}}
	\sum_{B\subset\Ni{i}: i\in B}
  \sa{i}{j}{B} \za{i}{j}{B,l},~\forall j\in \Kcal
\end{align}
where $L=\{0,\dots,k\}$ \BZ{is the index set \BZf{of} the $k+1$ segments}.
\DG{Basically, each $\za{i}{j}{B}$ defined in~\eqref{eq:LocalGlobal} is replicated $k+1$ times as $\Big( \za{i}{j}{B,0},\dots,\za{i}{j}{B,k} \Big)$ for the $k+1$ segments.}  There are altogether $O(kn)$ such $\za{i}{j}{B,l}$ variables.

\DG{We also introduce a set of local variables to take the place of the global $y$-variables.  Specifically, for every $l\in L$, $i\in N$, and $B\subset N_i$, let $\yA[i]{B,l}$ denote the total bandwidth}
assigned to AP $i$'s local pattern, $B\subset\Ni{i}$, within segment $l$.  
Then, in analogy to~\eqref{eq:opt:x<=y}, we have
\begin{align} \label{eq:Con2}
  \sum_{j\in\Ui{i}} \za{i}{j}{B,l} = \yA[i]{B,l},
  ~\forall l\in L,\; i\in N,\; B\subset\Ni{i}: i\in B.
\end{align}
Let $h\aol l$ denote the bandwidth \BZr{assigned to} segment $l$, where $\sum_{l\in\Lcal} h\aol l=1$.
Evidently, the total \DG{bandwidth \BZ{allocated to all local patterns of AP $i$} within segment $l$}
should equal 
$h^l$:
\begin{align} \label{eq:Con5}
  \sum_{B\subset \Ni{i}}\yA[i]{B,l}=h^l,~\forall i\in\Ncal,\, l\in\Lcal.
\end{align}
To enforce a one-to-one mapping of \DG{the $k+1$ active} patterns to 
segments, we add the \DG{following} 
constraint, \DG{which allows at most one local pattern to be activated in each interference cluster within each segment}:
\begin{align} \label{eq:Con6}
  \sum_{B\subset\Ni{i}} \left|\yA[i]{B,l}\right|_0\leq1,
  &~\forall i\in \Ncal ,\; l\in L
\end{align}
where \DG{the $\ell_0$-norm is defined as} $|x|_0=1$ if $x\neq 0$, and $|x|_0=0$ if $x=0$.

\DG{Collecting the preceding constraints}, 
we introduce the following problem formulation, referred to as~\ref{eq:Opt3}:
\begin{varsubequations}{P1} \label{eq:Opt3}
\begin{align}
  \maximize_{\br,\,\by,\,\bz} \;
  & u(\rj 1,\cdots,\rj k) \label{eq:Obj-Opt3}\\
  \subjectto \;
  & \rj j=\sum_{i\in\Aj{j}} \sum_{B\subset\Ni{i}: i\in B}
    \sa{i}{j}{B} 
		\sum_{l\in\Lcal} \za{i}{j}{B,l},
    \; \forall j\in \Kcal  \label{eq:Con1-Opt3}\\
  & \sum_{j\in\Ui{i}} \za{i}{j}{B,l} = \yA[i]{B,l},
    \; \forall l\in\Lcal, i\in \Ncal, B\subset\Ni{i}: i\in B \label{eq:Con2-Opt3}\\
  & \sum_{B\subset \Ni{i}:B\cap\Ni{m}=C} \yA[i]{B,l}
    = \sum_{B\subset \Ni{m}:B\cap\Ni{i}=C} \yA[m]{B,l}, \nonumber\\
  & \qquad\quad \forall l\in\Lcal,\, i,m\in\Ncal:N_i\cap N_m\neq\emptyset \nonumber\\
  & \qquad\quad \forall C\subset\Ni{i}\cap\Ni{m}:\, C\neq\emptyset \label{eq:Con3-Opt3}\\
  & \sum_{B\subset \Ni{i}} \yA[i]{B,l}= \BZr{h^l},
    \quad \forall l\in\Lcal,\, i\in\Ncal \label{eq:Con5-Opt3}\\
  & \sum_{B\subset\Ni{i}} \left|\yA[i]{B,l}\right|_0\leq1,
    \quad \forall l\in\Lcal,\, i\in\Ncal \label{eq:opt3:|y|_0}\\
  & \za{i}{j}{B,l}\geq 0,
    \;\; \forall l\in\Lcal, i\in\Ncal, j\in\Ui{i}, B\subset\Ni{i}\label{eq:Con9-Opt3}\\
  & \sum_{l\in\Lcal} \BZr{h^l}=1 \label{eq:Con7-Opt3} 
\end{align}%
\end{varsubequations}%
\DG{where~\eqref{eq:Con1-Opt3},~\eqref{eq:Con2-Opt3},~\eqref{eq:Con5-Opt3}, and~\eqref{eq:opt3:|y|_0} are identical to~\eqref{eq:Con1}--\eqref{eq:Con6}.  The additional constraint~\eqref{eq:Con3-Opt3} was introduced}
in~\cite{zhuang2017scalable} to ensure \DG{consistency of bandwidth allocations across overlapping clusters.  Basically,}
for 
every nonempty local pattern $C\subset\Ni{i}\cap\Ni{m}$, \DG{the total bandwidth allocated to $C$ in the interference cluster of AP $i$ must be identical to the total bandwidth allocated to $C$ in the interference cluster of AP $m$.  As an example, consider the network depicted}
in Fig.~\ref{fig:LocalPattern}. In interference cluster $\Ni{1}$, AP 2 transmits under pattern $\{2\}$ and $\{1,2\}$; while in interference cluster $\Ni{3}$, AP 2 transmits under pattern $\{2\}$ and $\{2,3\}$. \BZr{The overlapping pattern is $C=\{2\}$}. Since the 
\DG{same physical spectrum is allocated to AP 2 whether viewed in cluster $\Ni{1}$ or $\Ni{3}$,}
we have
$\yA[1]{\{2\},l} + \yA[1]{\{1,2\},l}
=\yA[3]{\{2\},l} + \yA[3]{\{2,3\},l}$.

\DG{We prove the following equivalence in Appendix.}

\begin{theorem} \label{thm:main}
  If~\eqref{eq:slocal} holds, then \BZr{the global formulation} \ref{eq:Opt} and \BZr{the local formulation}~\ref{eq:Opt3} are equivalent
  in the sense that they achieve the same maximum utility with the same optimal rate vectors.  In addition, given an optimal solution to~\ref{eq:Opt3}, the optimal global patterns to activate are
\begin{align}
\label{eq:OptPat}
    \Bl{l} &= \bigcup_{i\in\Ncal} \; \bigcup_{B\subset\Ni{i}: \yA[i]{B,l}>0} B, \quad l\in\Lcal
  \end{align}
  \BZa{and the corresponding solution to~\ref{eq:Opt} is given by
\begin{align}
\label{eq:OptVar}
\xa{i}{j}{A} &= \sum_{l:A=B^l} \za{i}{j}{\Bl{l},l}, \; \forall A\in N,\, i\in A,\, j\in\Ui{i}.
\end{align}
}
\end{theorem}

The detailed proof of Theorem~\ref{thm:main} is shown in the Appendix.

\section{Iterative $\ell_1$ Approximation}
\label{sec:l1}

The number of variables is reduced from $O(n2^n)$ in~\ref{eq:Opt} to $O(kn)$ in~\ref{eq:Opt3}. However, the $\ell_0$ norm constraint makes the problem non-convex and difficult to solve \DG{using standard solvers}.
\DG{In this section, we propose a} reweighted $\ell_1$ 
\DG{constraint in lieu of the $\ell_0$ norm constraint}
to 
\DG{obtain} a sparse solution.
This approach was previously proposed in~\cite{candes2008enhancing}
to address an $\ell_0$ norm in the optimization objective
and has been applied to various problems in the literature.
The basic idea here is to use the weighted $\ell_1$ norm, $\sum_i w\ioj i|y\ioj i|$ as a local approximation of the $\ell_0$ norm \BZ{$\sum_i|y_i|_0$} in constraint~\eqref{eq:opt3:|y|_0}, where $w\ioj i$ is updated in each iteration to be inversely proportional to the current $\ell_1$ norm, $|y\ioj i|$. The intuition is to discourage small nonzero entries with large weights.


\DG{Because} the $\ell_0$ norm constraints on different \BZa{segments} are related through~\eqref{eq:Con5-Opt3} and~\eqref{eq:Con7-Opt3},
the $\ell_1$ heuristic cannot be directly applied to~\ref{eq:Opt3}.
We develop an iterative algorithm based on the $\ell_1$ reweighted heuristic, where the weights depend on both the $\ell_1$ norm and the bandwidth allocated to each segment $h^l$. In each iteration of Algorithm~\ref{alg:l1}, we solve the following optimization problem:
\begin{varsubequations}{P2}
\label{eq:Opt6}
\begin{align}
\maximize_{\br,\,\by,\,\bz}~
& u(\rj 1,\cdots,\rj k) \label{eq:Obj-Opt6}\\
\subjectto~ & \sum_{B\subset\Ni{i}}\wi{i}{B,l}\yA[i]{B,l}\leq1,
~\forall l\in L,\; \forall i\in \Ncal\label{eq:Con5-Opt6}\\
&~\eqref{eq:Con1-Opt3},\; \eqref{eq:Con2-Opt3},\; \eqref{eq:Con3-Opt3},\; \eqref{eq:Con5-Opt3},\; 
\eqref{eq:Con9-Opt3},\; \eqref{eq:Con7-Opt3}.\nonumber
\end{align}
\end{varsubequations}%
The only difference from~\ref{eq:Opt3} is that we substitute the $\ell_0$ norm constraint~\eqref{eq:opt3:|y|_0} with the weighted sum~\eqref{eq:Con5-Opt6}\footnote{\BZf{Since $\yA[i]{B,l}\geq 0$ due to~\eqref{eq:Con2-Opt3} and~\eqref{eq:Con9-Opt3}, the $\ell_1$ norm is equivalent to the weighted sum.}}.

\begin{algorithm}
\caption{Iterative algorithm with reweighted $\ell_1$ approximation.}
\label{alg:l1}
\begin{algorithmic}[]
\INPUT {$(\sa{i}{j}B)_{j\in \Kcal , i\in \Aj{j}, B\subset N_i}$, and $(\lambda\ioj j)_{j\in \Kcal}$.}
\OUTPUT{The bandwidths allocated to the $k+1$ segments $(h^l)_{l\in L}$, the $k+1$ active patterns $(B^l)_{l\in L}$, and the spectrum allocated to link $i\to j$ on segment $l$, $(\ijA{\bar{x}}{i}{j}{l})_{j\in \Kcal ,i\in\Aj{j},l\in L}$}
\Init{Randomly choose $\wi{i}{B,l}\in(0,1),~i\in \Ncal ,~B\subset\Ni{i},~l\in L, ~\BZr{\mu\in (0,1)}$,.}
\item[Repeat]
    \State 1. Solve~\ref{eq:Opt6}, with the current weights $\bw$.
    \State 2. Update $\wi{i}{B,l}=\big(\yA[i]{B,l}+\mu h^l\big)^{-1},~\forall i\in\Ncal ,\; B\subset\Ni{i},\; l\in L$.
    \State 3. $t=t+1$.
\item[until] $(y_i^{B,l})_{i\in N,B\subset N_i,l\in L}$ converges or the maximum number of iteration is reached.
\Post {
\State 4. \DG{For all $l\in L$}, determine the optimal local patterns: 
\begin{align} \label{eq:DomPattern}
  \Bl[i]{l} = \arg\max_{B\in\Ni{i}}\, \yA[i]{B,l},  
\end{align}
and the corresponding global pattern:
\begin{align} \label{eq:UniPattern}
	\DG{B^l = \cup_{i\in\Ncal} \Bl[i]{l} . }
\end{align}
\State 5. \DG{For segment $l\in L$, the spectral efficiency of link $i\to j$ \BZf{becomes}:}
\begin{align}
\label{eq:SpeEff1}
\ijA{\bar{s}}{i}{j}{l}=
\begin{cases}
\sa{i}{j}{B^l\cap \Aj{j}},~\text{if } B^l\cap \Aj{j}\neq\emptyset,\\
0,~\text{otherwise}.
\end{cases}
\end{align}
\State 6. Optimize $(\ijA{\bar{x}}{i}{j}{l})_{j\in \Kcal ,i\in \Aj{j},l\in L}$ by solving the following optimization problem:}
\begin{varsubequations}{P3}
\label{eq:Opt7}
\begin{align}
  \maximize_{\bar{\bx},\;\bh}~
  & u(\rj 1,\cdots,\rj k) \label{eq:Obj-Opt7}\\
  \subjectto~
  & \rj j=\sum_{l\in L}\sum_{i\in \Aj{j}} \ijA{\bar{s}}{i}{j}{l} \ijA{\bar{x}}{i}{j}{l},
  	\quad \forall j\in K \label{eq:Con1-Opt7} \\
  & \sum_{j\in U_i} \ijA{\bar{x}}{i}{j}{l}= h^l,
	 	\quad \forall l\in K\label{eq:Con2-Opt7} \\
	& \sum_{l\in L} h^l =1 \label{eq:Con4-Opt7} \\
	& \ijA{\bar{x}}{i}{j}{l} \geq0,
		\quad \forall i\in \Ncal , j\in \Ui{i}, l\in L . \label{eq:Con3-Opt7}
\end{align}
\end{varsubequations}
\end{algorithmic}
\end{algorithm}

The iterative algorithm for solving~\ref{eq:Opt6} is described as Algorithm~\ref{alg:l1}. At the initial stage, a random initialization of the weights is used to introduce the necessary \BZf{asymmetry} in the first iteration. Otherwise, e.g., setting all $w_i^{B,l}=1$, the solution will be symmetric over all segments, which is not optimal in general. In each iteration, we solve~\ref{eq:Opt6} with the current weights $\bw$ to obtain the corresponding optimal $\bx,~\by,~\bz$ and \BZr{$\bh=[h^0,\cdots,h^{k}]$}. Then the weights are updated. The iteration terminates when the solution converges or the maximum number of iterations $t_{max}$\footnote{The maximum number of iterations is limited to 8 in our simulations.} is reached.

The weight update in Algorithm~\ref{alg:l1} is designed to approximate the $\ell_0$ norm (see~\cite{candes2008enhancing} and the reference therein). Because \BZf{we want to keep searching over all local patterns,}
$\mu h^l$ \BZf{is added} to the denominator, with $\mu\in(0,1)$. Note that we use a variable $\mu h^l$ unlike the fixed $\epsilon$ proposed in~\cite{candes2008enhancing}, \BZ{which adapts to the bandwidth $h^l$ change in different iterations}. Algorithm~\ref{alg:l1} simultaneously searches for the \BZr{optimal} pattern on each segment as well as the bandwidth allocated to it. \EWr{While the heuristic of iteratively approximating $\ell_0$ by reweighted $\ell_1$ norms lacks formal convergence guarantees, it performs well in practice as observed in~\cite{candes2008enhancing}.}

A consensus on a single pattern may not be reached for all segments when the iterations terminate. We \BZr{use~\eqref{eq:DomPattern} to} enforce a unique pattern on each segment by letting AP $i$ use \DG{a}
dominating pattern. The unified global pattern $B^l$ on segment $l$ is thus given by \BZ{the union of all the dominating patterns on segment $l$}, which is shown in~\eqref{eq:UniPattern}. Given the $k+1$ global patterns, we can determine the spectral efficiency for each link $(i\to j)$ \DG{under those patterns} provided by~\eqref{eq:SpeEff1}\footnote{\BZ{Since one pattern is used in each segment, the spectral efficiency is determined for each segment.}}. The bandwidth of all segments, $\bh=(h^l)_{l\in L}$ and the spectrum allocation over different links on each segment, $\bar{\bx}=\left( \ijA{\bar{x}}{i}{j}{l} \right)_{j\in\Kcal ,i\in\Aj{j},l\in L}$ are further optimized by solving the relatively simple problem~\ref{eq:Opt7}
\DG{in Algorithm~\ref{alg:l1}}
with $O(nk)$ variables.

One way of using Algorithm~\ref{alg:l1} to solve~\ref{eq:Opt3} \BZa{is} by passing the required parameters to a central controller to optimize the spectrum allocation and user association. The required parameters are channel information $\left(s^B_{i\to j}\right)_{j\in K,\; i\in A_j\; B\subset N_i}$ and traffic information $\left(\lambda_j\right)_{j\in K}$. These parameters are static and only need to be communicated once in each time period. Hence, the number of coefficients sent to the central controller is $O(k)$. The optimal solution obtained using Algorithm~\ref{alg:l1} can be represented by the optimal patterns $\left(B^l\right)_{l\in L}$, the optimal bandwidths of the segments $\left(h^l\right)_{l\in L}$, and the allocation variables $\left(\bar{x}^l_{i\to j}\right)_{j\in K,\; i\in A_j,\; l\in L}$, which is much less than the number of variables originally in~\ref{eq:Opt3}. Thus, the central controller only needs to feed back \BZa{$O(k^2)$} variables to inform the optimal allocation to the APs.

\section{A Distributed Algorithm based on ADMM}
\label{sec:ADMM}

ADMM originates from the augmented Lagrangian algorithm~\cite{ForGlo1983chapter3,
Gab1983chapter4}.
It solves a problem with decomposable objective by \DG{iteratively} solving small sub-problems and reconciling their results. The \mbox{ADMM} has been proved to be effective in solving many optimization problems that arise from ``big data''. ADMM based solutions can often be implemented in a distributed manner or make use of parallel computing to solve subproblems simultaneously. In the previous section, we introduced~\ref{eq:Opt6} as a convex approximation of~\ref{eq:Opt3}. Here we \BZa{show how to use an ADMM based algorithm to solve the convex problem~\ref{eq:Opt6} in a distributed way.}

We first present an equivalent formulation of\BZ{~\ref{eq:Opt6}}:
\begin{varsubequations}{P4}
\label{eq:Opt8}
\begin{align}
  \maximize_{\br,\,\bz,\,\by}~
  & u(\rj 1,\cdots,\rj k) \label{eq:Obj-Opt8}\\
  \subjectto~
  & \BZ{v\ioj{i,m}\aol{C,l}=\sum_{B\subset\Ni{i}:B\cap \Ni{m}=C}\yA[i]{B,l},}\nonumber\\
  &\DG{\quad \forall l\subset L, \; m,i\in N,\; C\subset N_i\cap N_m: C\neq\emptyset} \label{eq:Con1-ADMM}\\
  &\BZ{v\ioj{i,m}\aol{C,l}=v\ioj{m,i}\aol{C,l},} \nonumber\\
  &\DG{\quad \forall l\subset L, \; m,i\in N,\; C\subset N_i\cap N_m: C\ne\emptyset} \label{eq:Con2-ADMM}\\
  &\eqref{eq:Con1-Opt3},\; \eqref{eq:Con2-Opt3},\;\eqref{eq:Con5-Opt3},
  \;\eqref{eq:Con9-Opt3}, \;\eqref{eq:Con7-Opt3},\;\eqref{eq:Con5-Opt6}\nonumber.
\end{align}
\end{varsubequations}%
\BZa{The additional auxiliary variables $(v^{C,l}_{i,m})_{i,m\in N,\;l\in L, \; C\subset N_i\cap N_m:C\ne\emptyset }$ are for decomposing
the optimization problem into subproblems, which consist of only local variables.}

The augmented Lagrangian of~\ref{eq:Opt8} can be written as:
\begin{align} \label{eq:AugLag}
  \begin{split}
  &L(\bv,\by,\bz,\bh,\balpha,\bbeta,\bgamma)= \sum_{j\in K} u_j(\rj j)-\balpha^T(\by-\bF\bz) \\
  & -\frac{\rho}{2}(\by-\bF\bz)^T(\by-\bF\bz) -\bbeta^T(\mathbf{v}-\mathbf{G}\by)-\bgamma^T(\mathbf{H}\by-\bh) \\
  &-\frac{\rho}{2}(\mathbf{v}-\mathbf{G}\by)^T(\mathbf{v}-\mathbf{G}\by)  -\frac{\rho}{2}(\mathbf{H}\by-\bh)^T(\mathbf{H}\by-\bh),
\end{split}
\end{align}\noindent
where the rate variables $r_j$ \BZf{is} calculated by~\eqref{eq:Con1-Opt3}, and $\bv$, $\by$, $\bz$ and $\bh$ are the vectors containing all $v\ioj{i,m}\aol{C,l}$, $\yA[i]{B,l}$, $\za{i}{j}{B,l}$, and $h^l$ variables, respectively. \BZ{$\balpha$, $\bbeta$, and $\bgamma$ are the Lagrangian multipliers for the constraints~\eqref{eq:Con2-Opt3},~\eqref{eq:Con1-ADMM} and~\eqref{eq:Con5-Opt3}, respectively. In~\eqref{eq:AugLag}, the constraints~\eqref{eq:Con2-Opt3},~\eqref{eq:Con1-ADMM} and~\eqref{eq:Con5-Opt3} are written in vector form as $\by-\bF\bz=0$, $\bv-\bG\by=0$, and $\bH\by-\bh=0$, respectively.} The positive parameter $\rho$ controls the weight on the quadratic penalty terms, which also corresponds to the step size of the dual descent update in the ADMM based solution to be introduced. \BZa{We only consider the dual variables of the equality constraints~\eqref{eq:Con2-Opt3},~\eqref{eq:Con1-ADMM}, and~\eqref{eq:Con5-Opt3} in~\eqref{eq:AugLag}. The rest of the constraints are omitted here for simplicity, which will be considered when solving each subproblem.}

\begin{algorithm}
\caption{The ADMM based algorithm.}
\label{alg:ADMM}
\begin{algorithmic}[!t]
\INPUT{$\boldsymbol{\lambda}=[\lambda_1,\cdots,\lambda_k]^T,~\bs=\left(\sa{i}{j}{B}\right)_{j\in K, i\in A_j, B\subset N_i}$, \BZr{$\bw=\left(w^{B,l}_i\right)_{l\in L,i\in N,B\subset N_i}$}}
\OUTPUT{$\bv$, $\by$, $\bz$ and $\bh$}

\State Initialize $\bv_0$,~$\bz_0$,~$\bh_0$,~$\balpha_0$,~$\bbeta_0$,~$\bgamma_0$ and set $t=0$.
\While{$\by$, $\bz$ and $\bh$ have not converged and $t<t_{max}$}

    \State $\by_{t+1}=\underset{\by}{\arg\max}~ L(\bv_t,\by,\bz_t,\bh_t,\balpha_t,\bbeta_t,\bgamma_t)$
    \State $[\bz_{t+1};\bv_{t+1};\bh_{t+1}]=\underset{\bz,\bv,\bh}{\arg\max}~
    L(\bv,\by_{t+1},\bz,\bh,\balpha_t,\bbeta_t,\bgamma_t)$
    \State $\balpha_{t+1}=\balpha_t+\rho(\by_{t+1}-\bF\bz_{t+1})$
    \State $\bbeta_{t+1}=\bbeta_t+\rho(\bv_{t+1}-\bG\by_{t+1})$
    \State $\bgamma_{t+1}=\bgamma_t+\rho(\bH\by_{t+1}-\bh_{t+1})$
    \State $t=t+1$
\EndWhile
\end{algorithmic}
\end{algorithm}
An ADMM based iterative algorithm is shown in Algorithm~\ref{alg:ADMM} to solve~\ref{eq:Opt6}.
The algorithm takes any initialization. In each iteration, there are three steps to update the primal and dual variables. First, update $\by_{t+1}$ based on $\bv_t$, $\bz_t$, $\bh_t$ $\balpha_t$, $\bbeta_t$ and $\bgamma_t$ calculated from the previous iteration to minimize~\eqref{eq:AugLag}. Then, update $\bv_{t+1}$, $\bz_{t+1}$ and $\bh_{t+1}$ based on $\by_{t+1}$, $\balpha_t$, $\bbeta_t$ and $\bgamma_t$. The dual variables $\balpha_{t+1}$, $\bbeta_{t+1}$ and $\bgamma_{t+1}$ are updated at the end of each iteration with the newly updated primal variables.

We next explain the distributed computation and message sharing \BZa{used} in the preceding updates. The update of $\by$ can be decomposed into $n$ subproblems \BZr{associated with} each of the $n$ APs. Define the part of the augmented Lagrangian~\eqref{eq:AugLag} related to $\by$ as:
\BZr{
\begin{align}
  \label{eq:LagYZ}
  \begin{split}
    L_{[\by]}&(\bv_t,\by,\bz_t,\bh_t,\balpha_t,\bbeta_t,\bgamma_t) =\\
    & -\balpha_t^T(\by-\bF\bz_t)-\frac{\rho}{2}(\by-\bF\bz_t)^T(\by-\bF\bz_t) \\ 
    &-\bbeta_t^T(\bv_t-\bG\by)-\frac{\rho}{2}(\bv_t-\bG\by)^T(\bv_t-\bG\by)\\ 
    &-\bgamma_t^T(\bH\by-\bh_t)-\frac{\rho}{2}(\bH\by-\bh_t)^T(\bH\by-\bh_t)
  \end{split}
\end{align}}\noindent
where $\bv_t$, $\bx_t$, $\bh_t$, $\balpha_t$, $\bbeta_t$ and $\bgamma_t$ are obtained in the previous iteration. \BZr{The subproblem \BZa{associated with} AP $i$ is obtained by taking the part of~\eqref{eq:LagYZ} that depends on $\by_i=(y_i^{B,l})_{B\in N_i,l\in L}$:
\begin{varsubequations}{P\textsubscript{y}}
\label{eq:OptY}
\begin{align}
\begin{split}
\minimize_{\by_i}\;
&
\sum_{l\in L}\sum_{B\subset N_i:i\in B}(y^{B,l}_i-\sum_{j\in U_i}z^{B,l}_{i\to j} + \frac{\alpha^{B,l}_i}{\rho})^2\\
&+ \sum_{l\in L}\sum_{m\in N_i}\sum_{C\subset N_i\cap N_m: C\neq\emptyset}(v^{C,l}_{i,m}-\sum_{B\subset\Ni{i}:B\cap \Ni{m}=C}\yA[i]{B,l} + \frac{\beta^{C,l}_{i,m}}{\rho})^2\\
&+\sum_{l\in L}(\sum_{B\subset \Ni{i}} \yA[i]{B,l} - h^l +\frac{\gamma^l_i}{\rho})^2
\end{split}
\label{eq:OptY-Obj}\\
\subjectto\;
& \sum_{B\subset \Ni{i}}\wi{i}{B,l} \yA[i]{B,l}\leq 1, \;\; \forall l\in N,\label{eq:OptY-Con1}
\end{align}
\end{varsubequations}\noindent
\BZa{which requires} intermediate results: $\balpha_{i,t}=\left(\alpha^{B,l}_i\right)_{l\in L, B\subset N_i:i\in B}$, $\bbeta_{i,t}=\left(\beta^{C,l}_{i,m}\right)_{l\in L, m\in N_i, C\subset N_i\cap N_m: C\neq\emptyset}$, $\bgamma_{i,t}=\left(\gamma^l_i\right)_{l\in L}$, $\bh_t=\left(h^l\right)_{l\in L}$, $\bv_{i,t}=\left(v^{C,l}_{i,m}\right)_{l\in L, m\in N_i, C\subset N_i\cap N_m: C\neq\emptyset}$, and $\bz_{i,t}=\left(z^{B,l}_{i\to j}\right)_{l\in L, j\in U_i, B\subset N_i}$. We shall see that $\balpha_{i,t}$, $\bbeta_{i,t}$, $\bv_{i,t}$, and $\bgamma_{i,t}$ are updated locally at AP $i$, when introducing the corresponding subproblems. The variables in $\bz_{i,t}$ are updated at the devices in $U_i$. Hence, the information sharing due to \BZa{$\bz_{i,t}$} is within AP $i$'s local cluster $U_i$. Only $\bh$ are shared globally, which requires sharing $k+1$ real numbers in each iteration of Algorithm~\ref{alg:ADMM}.}


The update of $\bz$, $\bv$ and $\bh$ in Algorithm~\ref{alg:ADMM} can be \BZf{divided into the updates of $\bz$, $\bv$, and $\bh$, respectively}. The part of~\eqref{eq:AugLag} relates to $\bz$ is:
\begin{align}
\label{eq:LagX}
L_{[\bz]}(\bz,\by_{t+1},\balpha_t)=\sum_{j\in \Kcal } u_j(\rj j) -\balpha_t^T(\by_{t+1}-\bF\bz)-\frac{\rho}{2}|\by_{t+1}-\bF\bz|^2.
\end{align}
\BZr{
The subproblem \BZa{for} $\bz_j=(z_{i\to j}^{B,l})_{i\in A_j, B\subset N_i, l\in L}$ is given by:
\begin{varsubequations}{P\textsubscript{z}}
\label{eq:OptZ}
\begin{align}
  \maximize_{\bz_j}\;
  & u_j(r_j)- \sum_{l\in L}\sum_{i\in A_j}\sum_{B\subset N_i: i\in B}(y^{B,l}_i-\sum_{j\in U_i}z^{B,l}_{i\to j} + \alpha^{B,l}_i/\rho)^2\label{eq:OptZ-Obj}\\
  \subjectto~
  & \rj j=\sum_{i\in A_j}\sum_{B\subset N_i:i\in B}\sa{i}{j}{B} \sum_{l\in L}\za{i}{j}{B,l} \label{eq:OptZ-Con1}\\
  & \za{i}{j}{B,l}\geq0, ~\forall l\in L,\; i\in N,\; B\subset N_i.\label{eq:OptZ-Con2}
\end{align}
\end{varsubequations}\noindent
The constraints~\eqref{eq:OptZ-Con1} and~\eqref{eq:OptZ-Con2} contain only the parts of~\eqref{eq:Con1-Opt3} and~\eqref{eq:Con9-Opt3} for $\bz_j$, respectively. The message sharing includes $\by_{j,t}=\left(y^{B,l}_i\right)_{l\in L, i\in A_j, B\subset N_i:i\in B}$ and $\balpha_{j,t}=\left(\alpha^{B,l}_i\right)_{l\in L, i\in A_j, B\subset N_i:i\in B}$, which are updated and shared by the APs in device $j$'s local cluster $A_j$. The rest of the $z$ variables in~\eqref{eq:OptZ-Obj} are updated at devices that can be served by the same APs as device $j$, i.e., devices in $U_i$ such that $i\in A_j$. To emphasize the association of the variable $\bz_j$ and subproblem~\ref{eq:OptZ} to device $j$, we say the subproblem is solved at device $j$. However, in practice, the problem should be physically solved at an AP instead, e.g., the nearest AP to device $j$.}

The variable $\bv$ can be separately updated for each pair of $v\aol{C,l}\ioj{i,m}$ and $v\aol{C,l}\ioj{m,i}$ in~\eqref{eq:Con2-ADMM}. If we initialize with  $\beta\aol{C,l}\ioj{i,m}+\beta\aol{C,l}\ioj{m,i}=0$, it is easy to prove that the $\bv$ update is in the following closed form:
\begin{align}
v\aol{C,l}\ioj{i,m}&=v\aol{C,l}\ioj{m,i}\label{eq:Sol-W1}\\
&=\frac{1}{2}\left(\sum_{B:B\cap\Ncal_{m}=C,~B\subset\Ncal _{i}}\yA[i,t+1]{B,l}+ \sum_{B:B\cap\Ni{i}=C,~B\subset\Ni{m}}\yA[m,t+1]{B,l}\right).\label{eq:Sol-W2}
\end{align}
The computation in~\eqref{eq:Sol-W2} can be taken in parallel at different APs, \BZr{i.e., AP $i$ updates $\left(v\aol{C,l}\ioj{i,m}\right)_{l\in L,\; m\in N_i,\;C\in N_i\cap N_m:C\neq\emptyset}$. To compute~\eqref{eq:Sol-W2}, $\left(\yA[i,m]{B,l}\right)_{B: B\cap N_m =C, B\subset N_i}$ is available at AP $i$ and $\left(\yA[m,i]{B,l}\right)_{B: B\cap N_i =C, B\subset N_m}$ is updated at AP $i$'s \BZf{interferer,} AP $m$. Therefore, the message sharing is locally among AP $i$ and its interfering APs in $N_i$.}

The subproblem for solving $\bh$ is:
\begin{varsubequations}{P\textsubscript{h}}
\label{eq:OptH}
\begin{align}
  \minimize_{\bh}~
  &\sum_{i\in N}\sum_{l\in L}\left(\sum_{B\subset \Ni{i}} \yA[i]{B,l} - h^l +\gamma^l_i/\rho\right)^2\\
  \subjectto~
  & \DG{h^l\geq0, \;\; \forall l\in L} \\
  & \sum_{l\in L}h^l=1,
\end{align}
\end{varsubequations}\noindent
which can be easily solved with standard quadratic programming solver, since it only has $k+1$ variables. \BZr{To solve~\ref{eq:OptH}, each AP $i$ needs to pass $\left(\sum_{B\subset \Ni{i}} \yA[i]{B,l}\right)_{l\in L}$ and $\left(\gamma^l_i\right)_{l\in L}$, i.e., $2(k+1)$ real values, to a central controller for the computation.}

The update of dual variable $\balpha$ can be obtained in distributed manner at the $n$ APs:
\BZr{
\begin{align}
\label{eq:SubAlpha}
\balpha^{B,l}_{i}=\balpha^{B,l}_{i}+\rho\left(y^{B,l}_i-\sum_{j\in U_i}z^{B,l}_{i\to j}\right),~~\forall l\in L,\; B\subset N_i: i\in B,
\end{align}
where $\by_{i,t+1}=(y_i^{B,l})_{B\in N_i,l\in L}$ is updated locally at AP $i$ and $\bz_{i,t+1}=\left(z^{B,l}_{i\to j}\right)_{l\in L, j\in U_i, B\subset N_i}$ are provided by the devices in $U_i$.

Analogously, $\bbeta$ is updated at the $n$ APs:
\begin{align}
\label{eq:SubBeta}
\beta^{C,l}_{i,m}=\beta^{C,l}_{i,m}+\rho \left(v^{C,l}_{i,m}-\sum_{B\subset\Ni{i}:B\cap \Ni{m}=C}\yA[i]{B,l}\right),\\\nonumber~~\forall l\in L,\; m\in N_i,\; C\subset N_i\cap N_m : C\neq\emptyset,
\end{align}
where both $\by_{i,t+1}=(y_i^{B,l})_{B\in N_i,l\in L}$ and $\bv_{i,t+1}=\left(v^{C,l}_{i,m}\right)_{l\in L, m\in N_i, C\subset N_i\cap N_m: C\neq\emptyset}$ are updated at AP $i$ as explained above.

The $\bgamma$ update can also be performed at the $n$ APs:
\begin{align}
\label{eq:SubGamma}
\bgamma^l_{i}=\bgamma^l_{i}+\rho \left(\sum_{B\subset \Ni{i}} \yA[i]{B,l} - h^l\right),~~\forall l\in L,
\end{align}
where $\by_{t+1}^i=(y_i^{B,l})_{B\in N_i,l\in L}$ is \BZa{locally available}, and $\bh_{t+1}=\left(h^l\right)_{l\in L}$ are \BZf{broadcasted} to all APs.
}

\BZr{The updates in Algorithm~\ref{alg:ADMM} are divided into simple subproblems that can be solved in distributed manner for each AP or device. \BZa{Only local message sharing is required during this process,} except broadcasting the $k+1$ values $\left(h^l\right)_{l\in L}$ from the central controller to the $n$ APs, and receiving the $2(k+1)$ values $\left(\sum_{B\subset \Ni{i}} \yA[i]{B,l}\right)_{l\in L}$ and $\left(\gamma^l_i\right)_{l\in L}$ at the central controller from each AP $i$. The above distributed updates of Algorithm~\ref{alg:ADMM} can also be used for parallel computing by passing the required information to a cloud.}

The ADMM based Algorithm~\ref{alg:ADMM} solves the convex optimization step in each iteration of Algorithm~\ref{alg:l1}, which takes up most of the computationa~\BZa{cost}. Here, we also want to point out that the weights updates $\wi{i}{B,l}=\big(\yA[i]{B,l}+\mu h^l\big)^{-1}$ can also be carried out at each AP $i$. Even for the very simple post processing~\ref{eq:Opt7}, we can \BZa{derive a distributed solution based on ADMM}. Therefore, the entire Algorithm~\ref{alg:l1} can be implemented in a distributed manner. \EWr{ Since ADMM has been shown to converge to the optimal solution for any convex problem~\cite{boyd2011distributed}, the proposed method inherits the same convergence guarantee. \BZa{The suboptimal solution achieved by Algorithm~\ref{alg:l1} is due to approximating $\ell_0$ norm using reweighted $\ell_1$ norm.}}

\section{Numerical Results}
\label{sec:NumRes}
The solution obtained by Algorithm~\ref{alg:l1} is evaluated using numerical simulations. Unless specified \DG{otherwise}, the general assumptions are given as follows: Among the $n$ APs, one macro AP is located at the center of the area and $n-1$ pico APs are randomly dropped around it. The $k$ devices are assumed to be located on $k$ \DG{randomly chosen} lattice points in the network. Both distance based pathloss and \DG{shadowing} are considered \BZa{to obtain} the link gains. The common parameters used for all the simulations in this section are shown in Table~\ref{tab:par}.

\begin{table}
\caption{Parameter configurations.}
\label{tab:par}
\centering
\begin{tabular}{||c|c||}
\hline\hline
Parameter & Value/Function\\
\hline
pathloss exponent & 3\\
standard deviation of shadow fading & 3\\
macro transmit PSD & 5 $\mu$W/Hz\\
pico transmit PSD & 1 $\mu$W/Hz\\
noise PSD & $1\times 10^{-7}$ $\mu$W/Hz\\
total bandwidth & 20 MHz\\
average packet length & 1 Mb\\
\hline\hline
\end{tabular}
\end{table}

\subsection{Performance in Small Networks}
We compare the solutions to~\ref{eq:Opt} and~\ref{eq:Opt3} in a small network cluster with $n=10$ and $k=\DG{23}$. Since the number of variables is \DG{not too large in this case},
we solve both versions of~\ref{eq:Opt} with and without the local neighborhood approximation~\eqref{eq:slocal} using a standard convex optimization solver. The solution to~\ref{eq:Opt3} is obtained using \BZ{iterative $l_1$ reweighted algorithm} in Algorithm~\ref{alg:l1}. To solve~\ref{eq:Opt3}, we can either \BZa{compute the update in each iteration of Algorithm~\ref{alg:l1}} with a standard convex optimization solver or use the ADMM based distributed algorithm in Algorithm~\ref{alg:ADMM}. The local neighborhoods are constructed by considering the strongest four APs for each device. Two other simple schemes are also compared here. One is the full spectrum allocation with the maxRSRP association. The other is the optimal orthogonal allocation,\footnote{Both spectrum allocation and user association are optimized assuming each AP exclusively occupies a fraction of the spectrum.} i.e., the solution to~\ref{eq:Opt} under the additional assumption that only the
\DG{singleton}
patterns
$\{1\}, \{2\}, \dots, \{n\}$
are active.

\begin{figure} 
\centering
\includegraphics[width=\columnwidth]{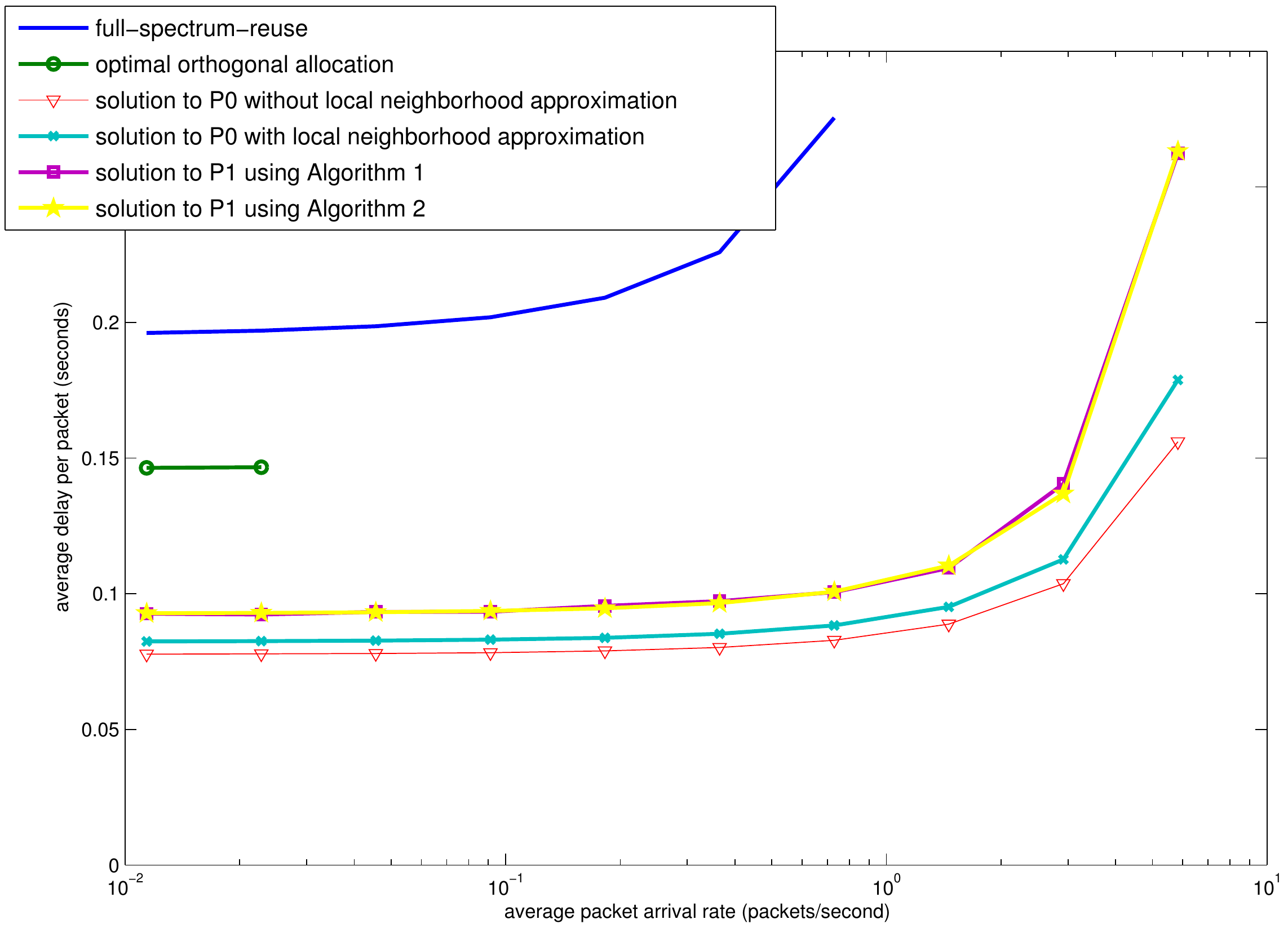}
\caption{Delay versus traffic arrival rate curves for a HetNet with $n=10$ and $k=23$.}
\label{fig:delay-N10K23}
\end{figure}

The delay versus traffic arrival rate curves are shown in Fig.~\ref{fig:delay-N10K23}. The \DG{rightmost} end of each curve represents the maximum arrival rate can be supported by the corresponding allocation scheme. The optimal orthogonal allocation (marked by circle marker) quickly becomes saturated, as the orthogonal spectrum allocation is very inefficient even only orthogonalizing over 10 APs. The full spectrum allocation with maxRSRP association (without any marker) achieves much higher throughput. However, the delay also increases \BZr{with respect to the optimal orthogonal allocation} as all APs cause interference to each other. The curves obtained by solving~\ref{eq:Opt} with and without local neighborhood approximation are very close, which indicates considering the four strongest interferers \BZr{and treating interference from remote APs as noise is accurate enough in the network setup}. The solution to~\ref{eq:Opt3} obtained by general convex solver and the ADMM based algorithm are almost on top of each other, which proves the validity of the ADMM based solution. Hence, in the subsequent results, we will only show the solution obtained using a standard convex solver in Algorithm~\ref{alg:l1}. The solutions to~\ref{eq:Opt3} achieve slightly longer delay than the solutions to~\ref{eq:Opt}. The maximum packet arrival rates that can be supported by the solutions to~\ref{eq:Opt} and~\ref{eq:Opt3} are the same. For all the simulations in this section, we limit the maximum number of iterations in Algorithm~\ref{alg:l1} to eight. The jointly optimized spectrum allocation and user association achieves substantial delay reduction as well as eight times throughput compared to the simple full spectrum allocation with maxRSRP association.

\begin{figure}
\centering
\subfloat[The solution to~\ref{eq:Opt}]{
\includegraphics[width=0.45\columnwidth]{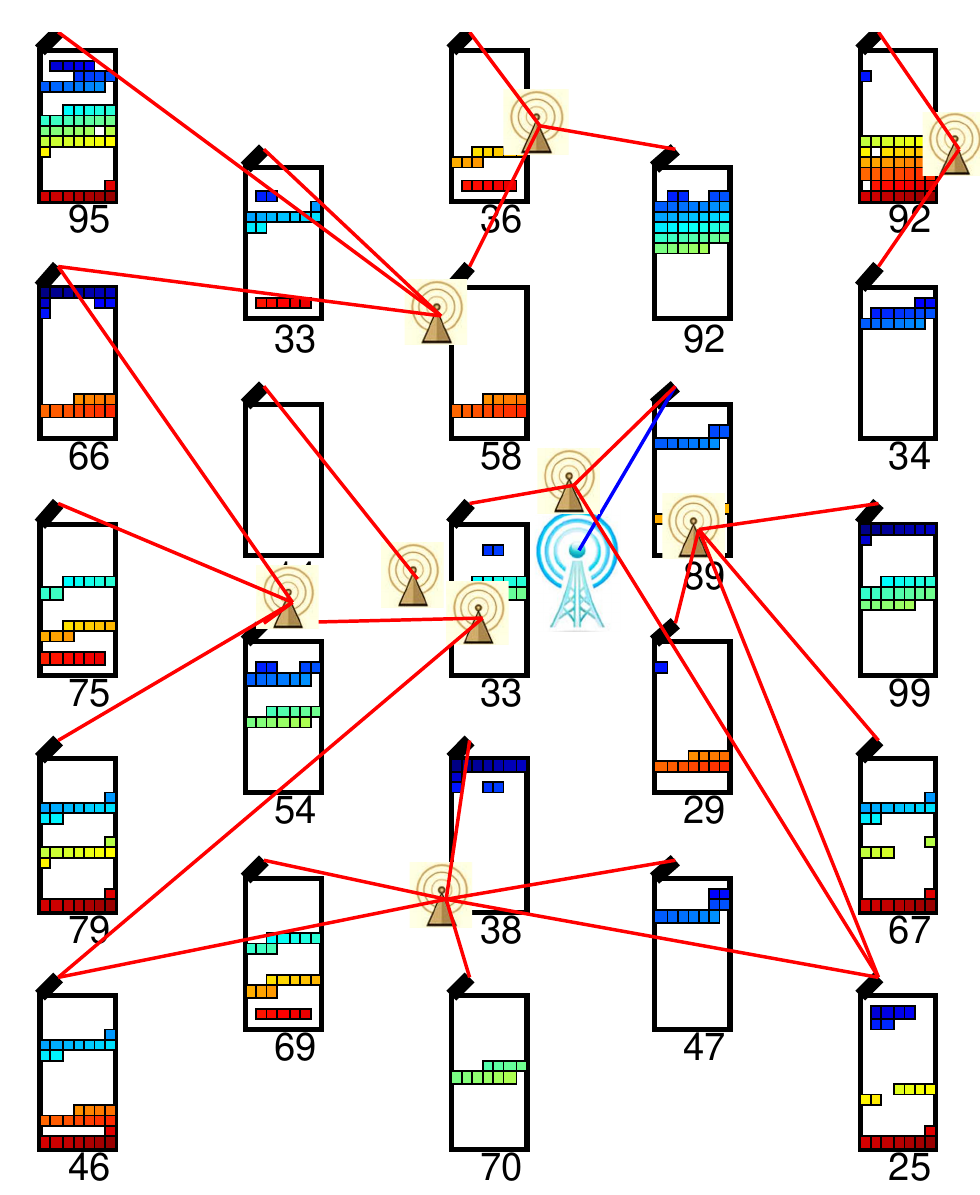}
\label{fig:Sol1}}
\subfloat[The solution to~\ref{eq:Opt3}]{
\includegraphics[width=0.45\columnwidth]{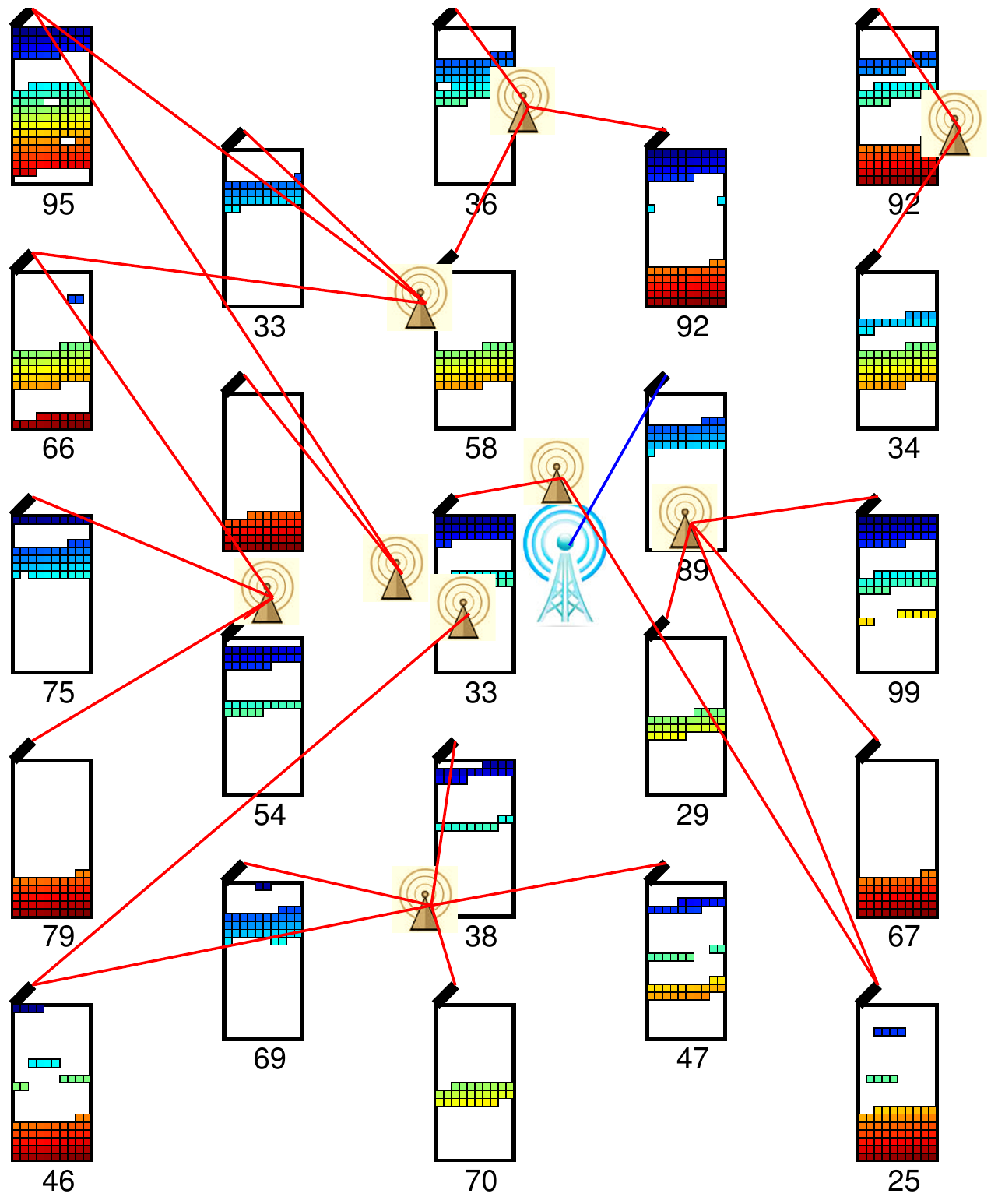}
\label{fig:Sol3}}
\caption{Proposed spectrum allocations and user associations at traffic arrival rate of 1.46 packets/second.}
\label{fig:TopoN10}
\end{figure}

The optimized spectrum allocations and user associations given by the solutions to~\ref{eq:Opt} (without local neighborhood approximation) and~\ref{eq:Opt3} are depicted in Fig.~\ref{fig:Sol1} and Fig.~\ref{fig:Sol3}, respectively. In Fig.~\ref{fig:TopoN10}, the macro and pico APs are represented by the bigger and smaller towers; each handset represents a device. If a device is associated to an AP, a solid line connects the corresponding AP and the device. The grid on each handset represents the spectrum used by the APs to serve \BZf{it}. The normalized traffic arrival rate (from 0 to 100) of each device is shown under each handset. The allocation achieved by the solution to~\ref{eq:Opt} with local neighborhood approximation is omitted, since it is almost identical to the solution to~\ref{eq:Opt} without local neighborhood approximation. The user association in Fig.~\ref{fig:Sol3} is close but not identical to the that in Fig.~\ref{fig:Sol1}. This is because the solution to~\ref{eq:Opt3} obtained using Algorithm~\ref{alg:l1} is an approximation to the \BZf{global optimum with local neighborhood approximation}. Take the device on the up left corner as an example, more APs transmit to it using a larger portion of the spectrum in Fig~\ref{fig:Sol3}~\BZr{compared to Fig.~\ref{fig:Sol1}}. \BZf{This also explains the delay difference between the solution to~\ref{eq:Opt} (with local neighborhood approximation) and the solution to~\ref{eq:Opt3} in Fig.~\ref{fig:delay-N10K23}.}

\subsection{Performance in Medium-size Networks}
\label{subsec:med}
\begin{figure}
\centering
\includegraphics[width=\columnwidth]{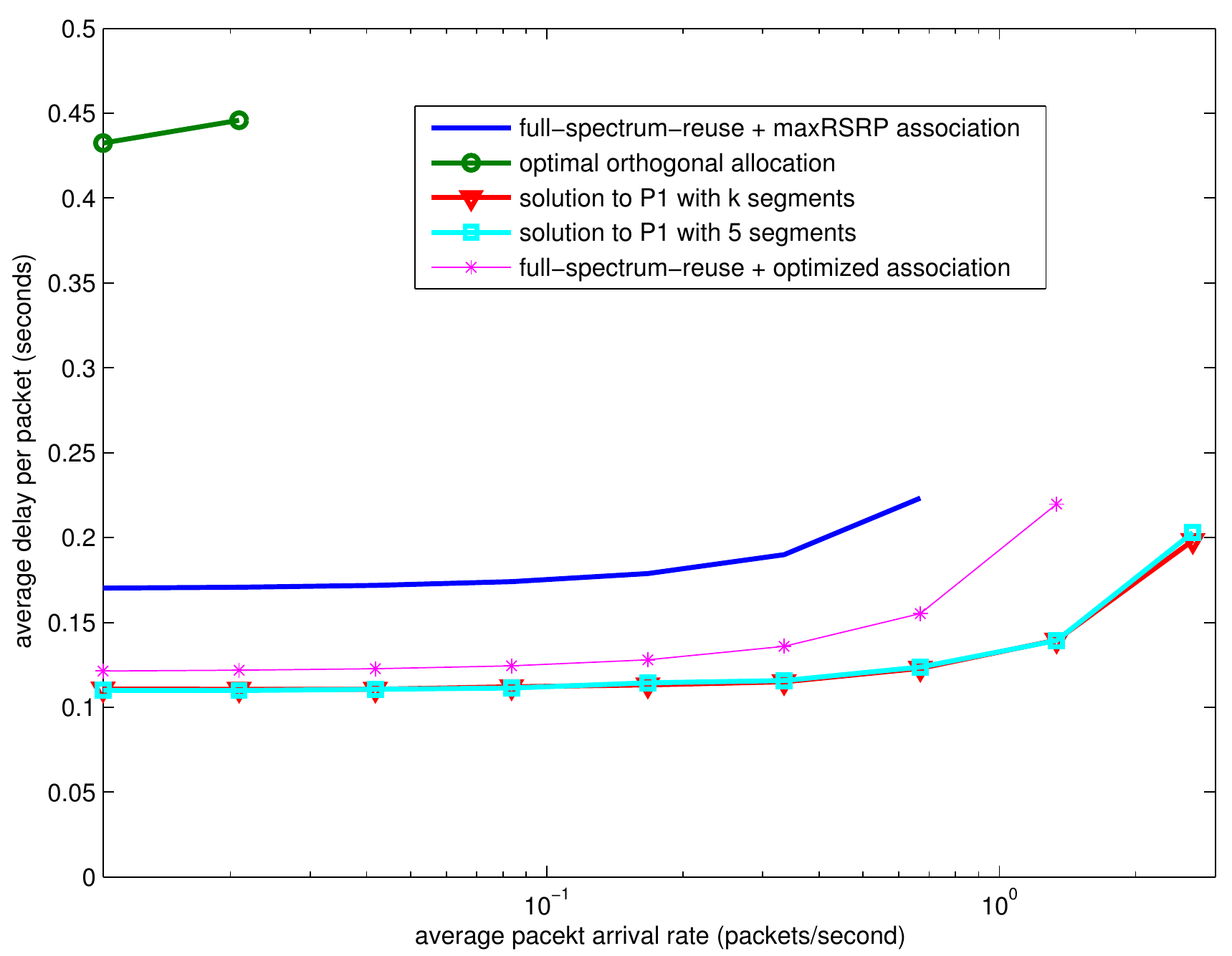}
\caption{Delay versus traffic arrival rate curves for a HetNet with $n=30$ and $k=46$.}
\label{fig:delay-N30K46}
\end{figure}

We present the performance comparison of different algorithms in a mid-size network with $n=30$ APs and $k=46$ devices, deployed on a $600\times600$ square meter area.~\ref{eq:Opt} becomes computationally prohibitive due to the $2^{30}$ global patterns. Hence we compare the solution to~\ref{eq:Opt3} with the simple maxRSRP association under the full-spectrum-reuse, the optimal orthogonal allocation and the optimal user association under \BZf{the full-spectrum-reuse}. A simplified version of~\ref{eq:Opt3} is also compared. Instead of using 46 segments, 5 segments are used, which constrain the solution to no more than five active patterns. The delay versus average traffic arrival rate curves obtained by the five different schemes are shown in Fig.~\ref{fig:delay-N30K46}. The optimal orthogonal allocation becomes even more inefficient. As the number of APs increases, each AP gets a smaller fraction of the entire spectrum on average. The solutions to~\ref{eq:Opt3} \BZr{using Aglorithm~\ref{alg:l1}} still achieves 4 times network throughput and substantial delay reduction compared with the full-spectrum-reuse with maxRSRP association. Interestingly, using 5 segments in~\ref{eq:Opt3} achieves almost the same performance as using 46 segments. This is because there are only seven active patterns in the solution to~\ref{eq:Opt3} with 46 segments. Many segments use the same active pattern in the solution. Optimizing user association under the full-spectrum-reuse also has superior performance over the full-spectrum-reuse with maxRSRP association. However, it can only support half of \BZf{the maximum traffic that can be supported by} the proposed solution.

\begin{figure}
\centering
\includegraphics[width=\columnwidth]{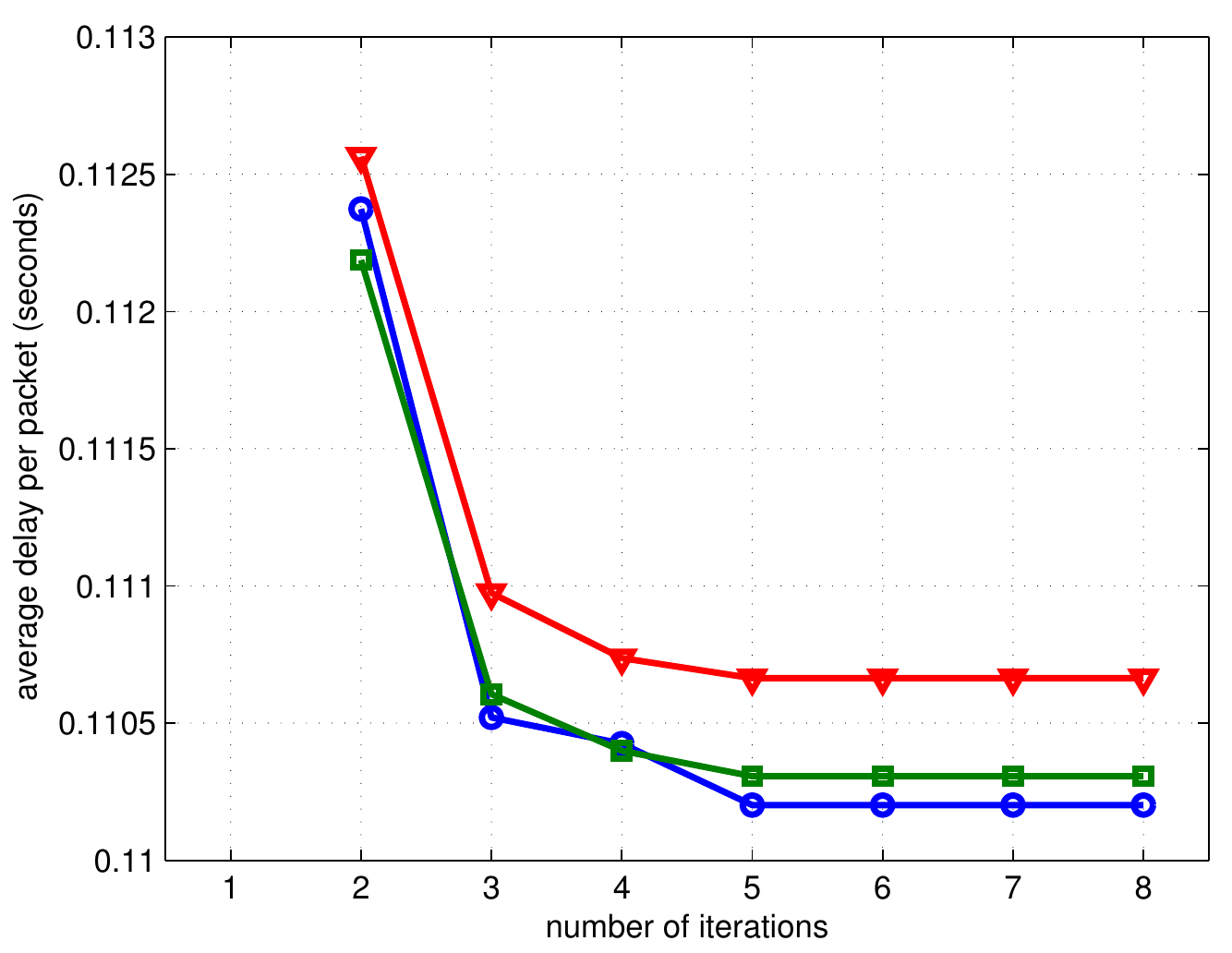}
\caption{Delay versus number of iterations in Algorithm~\ref{alg:l1}.}
\label{fig:DelayIter}
\end{figure}

\BZr{We also evaluate the convergence behavior of Algorithm~\ref{alg:l1} using this medium-size network. The average delay versus \BZa{iteration number} curves for the first three traffic loads in Fig.~\ref{fig:delay-N30K46} are shown in Fig.~\ref{fig:DelayIter}. To get a feasible spectrum allocation and the corresponding average delay at the end of each iteration, we perform post processing after every iteration. At the end of the first iteration, the dominating pattern on each segment given by~\eqref{eq:DomPattern} is still far from the optimal pattern. Hence,~\ref{eq:Opt7} has no feasible solution after the first iteration. That is why \BZf{no average delay values are shown} after the first iteration in Fig.~\ref{fig:DelayIter}. As the iteration continues, Algorithm~\ref{alg:l1} converges within five iterations under all three traffic loads. In fact, this kind of fast convergence has been observed throughout our simulations.}

\subsection{Performance in Large Networks}
\begin{figure}
\centering
\includegraphics[width=\columnwidth]{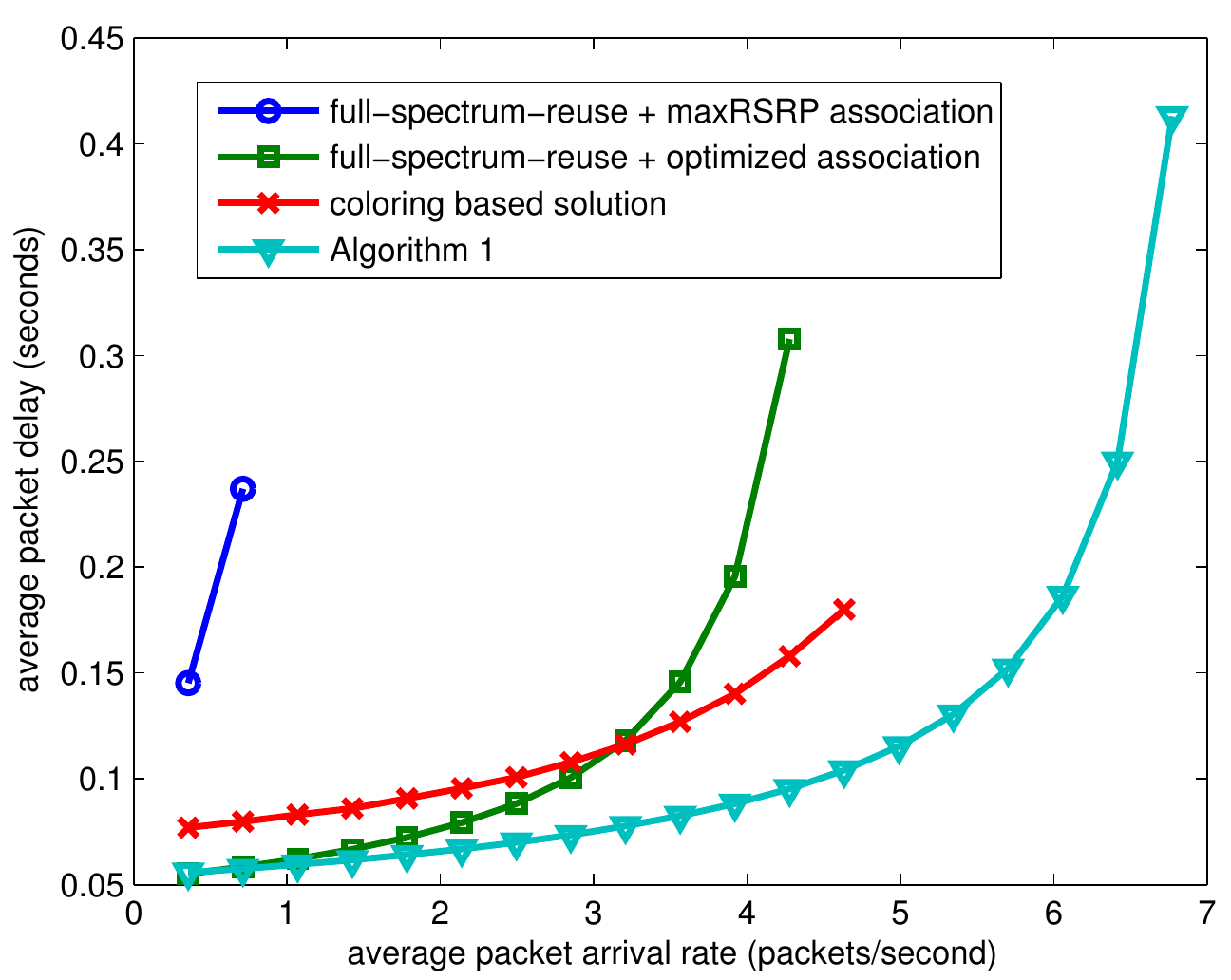}
\caption{Delay versus traffic arrival rate curves for a HetNet with $n=100$ and $k=200$.}
\label{N100K200}
\end{figure}

The performance of different allocation schemes are also compared in a large network with $n=100$ APs and $k=200$ devices, as shown in Fig.~\ref{N100K200}. The network is deployed on a $1250\times 1250$ square meter area. Here, we want to emphasize a `device' on slow timescales generally represents a class of service requests from different physical devices on fast timescales with the same QoS. Therefore, serving 200 devices on a slow timescale under heavy traffic \BZa{corresponds to} supporting thousands of users on fast timescales.
To 
\DG{ease computation}, \BZf{we reduce the sizes of local neighborhoods by constraining each device to be served by the three strongest APs. Under such constraint, the size of interference cluster $N_i$ is mostly between 5 to 8, in the large network.}

No optimal orthogonal allocation can support more than the lightest load \BZa{shown in Fig.~\ref{N100K200}.} Hence, we compare the full-spectrum-reuse with maxRSRP association, the full-spectrum-reuse with optimized association, the coloring based approach in~\cite{zhuang2017scalable}, and the solution to~\ref{eq:Opt3} obtained using Algorithm~\ref{alg:l1}. The coloring based approach suffers in this very large network due to the suboptimal solution based on various approximations, which is consistent with the observations in~\cite{zhuang2017scalable}. The coloring based solution even achieves higher average delay than the full-spectrum-reuse with optimized association in the light traffic regime. As the load increases, the coloring based approach outperforms the full-spectrum-reuse with optimized association. The proposed solution (Algorithm~\ref{alg:l1}) consistently outperforms all the other three schemes.
The throughput gain achieved by Algorithm~\ref{alg:l1} in this large network \BZr{is less than that in the medium-size network shown in section~\ref{subsec:med}}. This is mainly because we only consider the three strongest interferers, which compromises the benefit of interference management.

%

\section{Conclusion}
\label{sec:Con}
We have introduced a new aspect of future cellular networks with densely deployed APs through centralized radio resource management. Substantial performance improvement can be achieved by jointly optimizing spectrum allocation and user association across all APs on a slow timescale.
Advanced optimization techniques
\BZf{are used to solve the problem for large networks consisting of} many APs and devices. The proposed framework and scalable solution suggest a way \BZf{for centralized radio resource management on the metropolitan scale}. Power control and load dependent interference are not considered in the current problem formulation, which are future research directions.

\appendix[Proof of Theorem~\ref{thm:main}]
\label{sec:App}

We first present an equivalent formulation of~\ref{eq:Opt} under the local spectral efficiency definition in~\eqref{eq:slocal}:

\begin{proposition} \label{prop:Opt=Opt1}
  \DG{\ref{eq:Opt} is equivalent to the following problem,~\ref{eq:Opt1}:}
\begin{varsubequations}{P5}
\label{eq:Opt1}
\begin{align}
  \maximize_{\br,\,\by,\,\bz}~
  & u(\rj 1,\cdots,\rj k) \\
  \subjectto~
  & \rj j =\sum_{i\in \Aj{j}}\sum_{B\subset \Ni{i}:i\in B}\sa{i}{j}{B} \za{i}{j}{B}, \quad
  \forall j\in\Kcal \label{eq:Con2-Opt1}\\
  & \za{i}{j}{B}=\sum_{A\subset\Ncal:\, A\cap\Ni{i}=B}\xa{i}{j}{A},\quad \forall i\in\Ncal ,\; j\in \Ui{i},\; B\subset\Ni{i}\label{eq:Con1-Opt1}\\
  &\sum_{j\in U_i }\xa{i}{j}{A} = \yA{A}, \quad\forall A\subset\Ncal ,\; i\in  A\label{eq:Con4-Opt1}\\
  &\sum_{A\subset \Ncal }\yA{A} = 1  \\ 
  &~\xa{i}{j}{A}\geq0, \quad\forall A\subset \Ncal ,\; i\in A,\; j\in\Kcal. 
\end{align}
\end{varsubequations}%
\end{proposition}

\begin{IEEEproof}
To show the equivalence, we only need to prove that~\eqref{eq:opt:r=x} is equivalent to the combination of~\eqref{eq:Con1-Opt1} and~\eqref{eq:Con2-Opt1}. Under the local neighborhood assumption~\eqref{eq:slocal}, this is exactly what we have derived in~\eqref{eq:LocalGlobal} and~\eqref{eq:riA}--\eqref{eq:eq4}.

\end{IEEEproof}
We next prove~\ref{eq:Opt3} is equivalent to~\ref{eq:Opt1} by introducing two more intermediate equivalent problems.
\begin{proposition} \label{prop:Opt1=Opt4}
  \ref{eq:Opt1} \DG{is equivalent to the following problem,}~\ref{eq:Opt4}: 
\begin{varsubequations}{P6}
\label{eq:Opt4}
\begin{align}
  \maximize_{\br,\,\bx,\,\by,\,\bz}~
  & u(\rj 1,\cdots,\rj k) \\
  \subjectto~
  &\rj j=\sum_{i\in\Aj{j}}\sum_{B\subset\Ni{i}} \sa{i}{j}{B}\sum_{l\in L}\za{i}{j}{B,l}, \;\; \forall j\in \Kcal  \label{eq:Con1-Opt4}\\
  &\za{i}{j}{B,l}=\sum_{A\subset N:A\cap\Ni{i}=B}\xa{i}{j}{A,l}, \nonumber\\
  &\qquad\qquad \forall i\in\Ncal ,\; j\in\Ui{i},\; B\subset\Ni{i},\; l\in L \label{eq:Con2-Opt4}\\
&\sum_{j\in U_i }\xa{i}{j}{A,l}=\yA{A,l}, \quad \forall A\subset \Ncal ,\;i\in A\; l\in L\label{eq:Con3-Opt4}\\
&\sum_{A\subset \Ncal }\yA{A,l}= h^l, \quad \forall l\in L \label{eq:Con5-Opt4}\\
&\sum_{A\in\Ncal }|\yA{A,l}|_0\leq 1, \quad ~\forall l\in L\label{eq:Con6-Opt4}\\
&\sum_{l\in L}h^l=1 \label{eq:Con7-Opt4}\\
&\xa{i}{j}{A,l}\geq 0, \quad \forall A\subset \Ncal ,\; i\in A,\; j\in \Ui{i},\;  l\in L. \label{eq:Con8-Opt4}
\end{align}
\end{varsubequations}%
\end{proposition}

\begin{IEEEproof} 
\ref{eq:Opt4} can be considered as first reformulating~\ref{eq:Opt1} by having $k+1$ constituents of the $\bx$, $\by$, $\bz$ variables for the $k+1$ spectrum segments, and then adding the cardinality constraint~\eqref{eq:Con6-Opt4} to guarantee \BZa{one-to-one} mapping between active patterns and spectrum segments.
\BZ{We show \DG{that every} optimal solution to~\ref{eq:Opt4}
\DG{corresponds to an optimal} solution to~\ref{eq:Opt1}, in the sense that they achieve the \DG{same rate vector $\br$ as well as the} same utility.

\DG{First, given an optimal} solution to~\ref{eq:Opt4}, we can combine the variables of the $k+1$ segments into a feasible solution to~\ref{eq:Opt1}:}
\begin{align}
&z^B_{i\to j}=\sum_{l\in L}z^{B,l}_{i\to j}, ~\forall i\in N,\;  j\in U_i,\; B\subset N_i\label{eq:k+1to1z}\\
&x^A_{i\to j}=\sum_{l\in L}x^{A,l}_{i\to j}, ~\forall A\subset N,\; i\in A,\; j\in U_i\label{eq:k+1to1x}\\
&y^A=\sum_{l\in L}y^{A,l}, ~\forall A\subset N.\label{eq:k+1to1y}\noindent
\end{align}
It is easy to check the variables $\bx$, $\by$ and $\bz$ constructed according to~\eqref{eq:k+1to1z},~\eqref{eq:k+1to1x}, and~\eqref{eq:k+1to1y} satisfy all the constraints in~\ref{eq:Opt1}. \BZ{According to~\eqref{eq:Con2-Opt1} and~\eqref{eq:Con1-Opt4}, the two solutions also achieve the same rate vector $\br$, \DG{hence also} the same utility.}

\BZ{It remains to show that an optimal solution to~\ref{eq:Opt1}
\DG{corresponds} to a feasible
solution to~\ref{eq:Opt4}}. \BZ{According to {\it Theorem}~\ref{thm:Spec} and {\it Proposition}~\ref{prop:Opt=Opt1}, there exists an optimal solution to~\ref{eq:Opt1} that activates at most $k+1$ global patterns $\yA{A}$. Suppose there are $k'\le k+1$ active patterns in such an optimal solution, which is denoted by $A_1,\cdots,A_{k'}$.}  \DG{We form a feasible solution to~\ref{eq:Opt4} as:}
\begin{align}
h^l&=\left\{
\begin{array}{ll}
y^{A_l} & \text{if} ~l\in\{1,\cdots,k'\}\\
0       & \text{otherwise}
\end{array}
\right.\label{eq:1tok+1h}\\
\yA{A,l}&=\left\{
\begin{array}{ll}
y^{A_l}  & \text{if} ~l\in\{1,\cdots,k'\},\; A=A_l\\
0        & \text{otherwise}
\end{array}
\right.\label{eq:1tok+1y}\\
\xa{i}{j}{A,l}&=\left\{
\begin{array}{ll}
x_{i\to j}^{A_l} & \text{if} ~l\in\{1,\cdots,k'\},\;A=A_l,\; i\in A,\; j\in U_i\\
0                & \text{otherwise}.
\end{array}
\right.\label{eq:1tok+1x}
\end{align}
\BZ{After obtaining $\bx$ by~\eqref{eq:1tok+1x}, $\bz$ can be calculated according to~\eqref{eq:Con2-Opt4}.} We essentially assign the $k'$ active patterns to the first $k'$ segments; and set the bandwidths of the rest of the segments to zero, i.e., $h^l=0,\;l=k'+1,\cdots,k+1$.
It is easy to verify that the solution to~\ref{eq:Opt4} \BZ{formed by~\eqref{eq:1tok+1h}--\eqref{eq:1tok+1x}}
satisfies all the constrains in~\ref{eq:Opt4} and achieve the same rate tuple $\br$ as the optimal solution to~\ref{eq:Opt1}.

Therefore,~\ref{eq:Opt1} and~\ref{eq:Opt4} \DG{are equivalent}.
\end{IEEEproof}

\begin{proposition} \label{prop:Opt4=Opt5}
  \ref{eq:Opt4} is equivalent to the following problem,~\ref{eq:Opt5}:
\begin{varsubequations}{P7}
\label{eq:Opt5}
\begin{align}
  \maximize_{\br,\,\by,\,\bz}~
  & u(\rj 1,\cdots,\rj k)
  \label{eq:Obj-Opt5}\\
  \subjectto~
  &\rj j=\sum_{i\in\Aj{j}}\sum_{B\subset\Ni{i}} \sa{i}{j}{B} \sum_{l\in L}\za{i}{j}{B,l}, \; \forall j\in \Kcal  \label{eq:Con1-Opt5}\\
  &\sum_{j\in\Ui{i}}\za{i}{j}{B,l}= \sum_{A\subset\Ncal : A\cap\Ni{i} =B}\yA{A,l}, \;\forall l\in L\; i\in N,\; B\subset\Ni{i} \label{eq:Con2-Opt5}\\
  &\sum_{A\subset \Ncal }\yA{A,l}= h^l, \quad \forall l\in L \label{eq:Con5-Opt5}\\
&\sum_{A\subset\Ncal }|\yA{A,l}|_0\leq1, \quad \forall l\in L \label{eq:Con6-Opt5}\\
&\yA{A,l}\geq0, \quad \forall A\subset\Ncal, \; l\in L \label{eq:Con7-Opt5}\\
&\za{i}{j}{B,l}\geq 0, \quad~\forall i\in\Ncal \; j\in\Ui{i},\; B\subset\Ni{i},\;l\in L\label{eq:Con9-Opt5}\\
&\sum_{l\in L}h^l=1. \label{eq:Con8-Opt5}
\end{align}
\end{varsubequations}%
\end{proposition}

\begin{IEEEproof}
\label{prf:Opt4=Opt5}
Despite the difference between the constraints~\eqref{eq:Con2-Opt4}, ~\eqref{eq:Con3-Opt4} and the constraint~\eqref{eq:Con2-Opt5}, ~\ref{eq:Opt4} and~\ref{eq:Opt5} are equivalent.\footnote{In~\ref{eq:Opt5}, the constraint~\eqref{eq:Con7-Opt5} can be derived from~\eqref{eq:Con3-Opt4} and~\eqref{eq:Con8-Opt4}.} Essentially,~\ref{eq:Opt5} can be considered as removing \BZa{the $\bx$ variable} from~\ref{eq:Opt4} and directly relating $\bz$ and $\by$ through~\eqref{eq:Con2-Opt5}. First, we can prove that~\eqref{eq:Con2-Opt4} and~\eqref{eq:Con3-Opt4} imply~\eqref{eq:Con2-Opt5} by:
\begin{align}
\sum_{j\in U_i}z^{B,l}_{i\to j}&=\sum_{j\in U_i}\sum_{A\subset \Ncal:A\cap N_i=B}x^{A,l}_{i\to j}\label{eq:step1}\\
&=\sum_{A\subset\Ncal:A\cap N_i=B}y^{A,l},\label{eq:step2}\noindent
\end{align}
where~\eqref{eq:step1} is due to~\eqref{eq:Con2-Opt4}; and~\eqref{eq:step2} is due to~\eqref{eq:Con3-Opt4}. \BZ{This suggests any solution to~\ref{eq:Opt4} is also a feasible solution to~\ref{eq:Opt5}.}

\BZ{Next, we prove that any solution to~\ref{eq:Opt5} also corresponds to a feasible solution to~\ref{eq:Opt4}. Due to the cardinality constraint~\eqref{eq:Con6-Opt5}, a feasible solution to~\ref{eq:Opt5} has one active global pattern per segment.}
\BZ{We denote the active pattern in segment $l$ as $B^l$, which is defined in~\eqref{eq:OptPat}. The $\by$ and $\bz$ variables in a feasible solution to~\ref{eq:Opt5} will satisfy all constraints in~\ref{eq:Opt4} except~\eqref{eq:Con2-Opt4} and~\eqref{eq:Con3-Opt4}.} \BZ{We then construct  $x^{A,l}_{i\to j}$ in the corresponding solution to~\ref{eq:Opt4} as:}
\DG{\begin{align}
\label{eq:x-z}
x^{A,l}_{i\to j}=
\begin{cases}
  z^{B_l\cap N_i,l}_{i\to j},  &\text{if } A=B_l\\
  0,                           &\text{otherwise}
\end{cases}
\end{align}
for all $i\in N$, $j\in U_i$, and $A\subset B$.}
\BZa{To see that $\bx$ constructed according to~\eqref{eq:x-z} satisfies~\eqref{eq:Con2-Opt4}, we only need to prove for the nonnegative variables.} The only active local pattern in cluster $N_i$ on segment $l$ is given by $B=B^l\cap N_i$; and the only active global pattern on segment $l$ is $A=B^l$. Hence~\eqref{eq:Con2-Opt4} is satisfied due to~\eqref{eq:x-z}. Given~\eqref{eq:Con2-Opt5} and~\eqref{eq:x-z},~\eqref{eq:Con3-Opt4} directly holds. Therefore, the converse is proved. Hence,~\ref{eq:Opt4} and~\ref{eq:Opt5} are equivalent.

\end{IEEEproof}

\begin{proposition} \label{prop:Opt3=Opt5}
  \ref{eq:Opt5} is equivalent to \ref{eq:Opt3}.
\end{proposition}

\begin{IEEEproof}
\label{prf:Opt3=Opt5}
The difference between~\ref{eq:Opt3} and~\ref{eq:Opt5} is only in the $\by$ variables, where $\yA[i]{B,l}$ in~\ref{eq:Opt3} is associated with the local pattern $B\subset\Ni{i}$, and $\yA{A,l}$ in~\ref{eq:Opt5} is associated with global pattern $A\subset\Ncal $. The objectives of~\ref{eq:Opt3} and~\ref{eq:Opt5} are identical. The other constraints are equivalent except for constraints~\eqref{eq:Con2-Opt3}-\eqref{eq:opt3:|y|_0} and~\eqref{eq:Con2-Opt5}-\eqref{eq:Con6-Opt5}. It remains to show that the two optimization problems share \DG{a common} optimal solution. Given the variables $y^{A,l}$ in~\ref{eq:Opt5}, the variables $y^{B,l}_i$ in~\ref{eq:Opt3} can be constructed by:
\begin{align}
\label{eq:Opt3->Opt5}
\yA[i]{B,l}=\sum_{A\subset\Ncal :A\cap\Ni{i}=B}\yA{A,l}, ~\forall i\in\Ncal ,\; B\subset\Ni{i},\; l\in L.
\end{align}
Hence~\eqref{eq:Con2-Opt3} becomes~\eqref{eq:Con2-Opt5}.
Suppose all constraints in~\ref{eq:Opt5} hold. \BZ{Then~\eqref{eq:opt3:|y|_0} is given by~\eqref{eq:Opt3->Opt5} and~\eqref{eq:Con6-Opt5}. For any nonempty set $C$ such that $C\subset N_i\cap N_m$, we also have:}
\begin{align}
\label{eq:y_eq}
\sum_{B\subset\Ni{i}:B\cap\Ni{m}=C}\yA[i]{B,l}&=\sum_{B\subset\Ni{i}:B\cap\Ni{m}=C}\sum_{A\subset\Ncal :A\cap\Ni{i}=B}\yA{A,l}\\
&=\sum_{A\subset\Ncal :A\cap\Ni{i}\cap\Ni{m}=C}\yA{A,l}\\
&=\sum_{B\subset N_m: B\cap N_i=C}\sum_{A\subset N:A\cap N_m=B}y^{A,l}\\
&=\sum_{B\subset N_m:B\cap N_i =C}y^{B,l}_m\noindent
\end{align}
which implies~\eqref{eq:Con3-Opt3}. Moreover,~\eqref{eq:Opt3->Opt5} also suggests:
\begin{align}
\label{eq:sumOpt3Opt5}
\sum_{B\subset\Ni{i}}\yA[i]{B,l}&=\sum_{B\subset\Ni{i}}\sum_{A\subset\Ncal :A\cap\Ni{i}=B}\yA{A,l}\\
&=\sum_{A\subset\Ncal }\yA{A,l} \\
&= h^l .
\end{align}
Hence~\eqref{eq:Con5-Opt3} is satisfied. Similarly,~\eqref{eq:opt3:|y|_0} is established as:
\begin{align}
\label{eq:L0Opt3Opt5}
\sum_{B\subset\Ni{i}}|\yA[i]{B,l}|_0 &=\sum_{B\subset\Ni{i}}\left|\sum_{A\subset\Ncal :A\cap\Ni{i}=B}\yA{A,l}\right|_0\\
&\leq\sum_{B\subset\Ni{i}}\sum_{A\subset\Ncal :A\cap\Ni{i}=B}|\yA{A,l}|_0\\
&\leq\sum_{A\subset\Ncal }|\yA{A,l}|_0\\
&\leq1.
\end{align}
Therefore, any solution in the feasible set of~\ref{eq:Opt5} will be a feasible solution to~\ref{eq:Opt3}, \BZ{i.e., the feasible set of~\ref{eq:Opt3} includes that of~\ref{eq:Opt5}.}

It remains to show the optimal solution to~\ref{eq:Opt3} belongs to the feasible set of~\ref{eq:Opt5}. The key is to reconstruct global variables $\yA{A,l}$ from the local variables $\yA[i]{B,l}$. Let us focus on a specific segment $l\in L$. Constraint~\eqref{eq:opt3:|y|_0} dictates that there is at most one active local pattern in each local cluster $N_i$. That is to say we can identify one active pattern $\Bl[i]{l}\subset\Ni{i}$ for every AP $i\in\Ncal $, such that $\yA[i]{B,l}=0,~\forall B\neq \Bl[i]{l}$. \BZ{In fact all these active patterns from each AP's local cluster will be assigned the same bandwidth on the same segment $l$, i.e., $y^{B^l_i,l}_{i}=y^{B^l_m,l}_{m},~\forall i,m\in N,\;\forall l\in L$. If the two local patterns satisfy $B^l_i\cap N_m= B^l_m\cap N_i\ne \emptyset$, we must have $y^{B^l_i,l}_{i}=y^{B^l_m,l}_{m}$ according to~\eqref{eq:Con3-Opt3}. Therefore, the APs on a segment $l$ are divided into groups of interfering APs. In each group, all the APs will assign the same bandwidth to its local active pattern. We can also see the bandwidths assigned to different groups are all equal to $h^l$ in an optimal solution. This is because, if one group assign less than $h^l$ bandwidth, we can proportionally scale up the nonzero $\by$ and $\bz$ variables in this group until the $\by$ variables reach $h^l$. After such update, the solution is still feasible and the utility is improved.
Hence all local active patterns on each segment $l$ corresponds to a common global pattern:}
\begin{align}
\label{eq:GlobalPattern}
A_l = \cup_{i\in\Ncal }\Bl[i]{l} \quad \forall l\in L.
\end{align}
The bandwidth assigned to this active global pattern is $h^l$, which suggests the global variables are given by:
\begin{align}
y^{A,l}=\left\{
\begin{array}{lr}
h^l &\text{if}~A=A_l\\
0 &\text{otherwise}
\end{array}
\right.\label{eq:GlobalY}
\end{align}
It is easy to verify $\left(y^{A,l}\right)_{A\subset\Ncal ,l\in L}$ satisfies~\eqref{eq:Con2-Opt5}-\eqref{eq:Con7-Opt5}. Hence, the optimal solution to~\ref{eq:Opt3} is in the feasible set of~\ref{eq:Opt5}. The equivalence is therefore established.
\end{IEEEproof}

\DG{Propositions~\ref{prop:Opt=Opt1}
--\ref{prop:Opt3=Opt5} imply that~\ref{eq:Opt},~\ref{eq:Opt1},~\ref{eq:Opt4},~\ref{eq:Opt5}, and~\ref{eq:Opt3} are all equivalent.
Hence the proof of Theorem~\ref{thm:main}.}

%
%

\section*{Acknowledgment}

The authors thank Dr.~Weimin Xiao and Dr.~Jialing Liu for \DG{stimulating} discussions.


\begin{IEEEbiography}[{\includegraphics[width=1in,height=1.25in,clip,keepaspectratio]{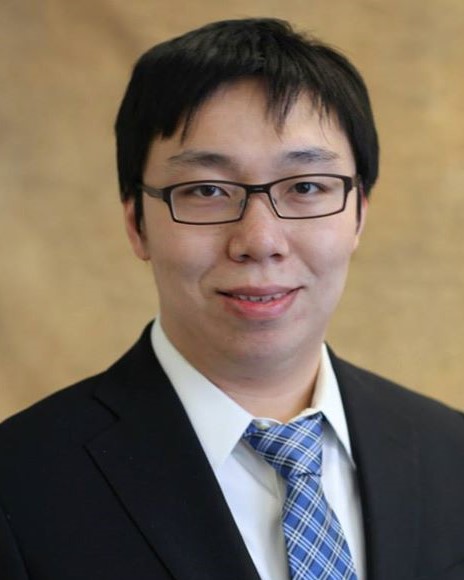}}]{Binnan Zhuang}
received his B.S. degree from Electronic Engineering Department of Tsinghua University, Beijing, China, in 2009, the M.S.\ and Ph.D.\ degrees in electrical engineering from Northwestern University, Evanston, IL, USA. in 2010 and 2015, respectively. He is currently working as a staff engineer in the System on Chip (SoC) Lab of Samsung Semiconductor Inc. in San Diego, CA, USA. His research interests include wireless communications, communication network and network optimization. His current research at Samsung focus on computer vision and communication.
\end{IEEEbiography}

\begin{IEEEbiography}[{\includegraphics[width=1in,height=1.25in,clip,keepaspectratio]{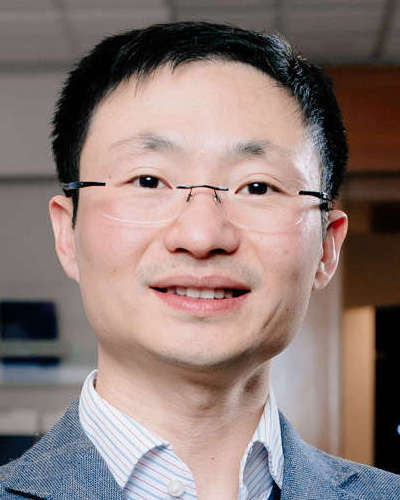}}]{Dongning Guo}
(S'97-M'05-SM'11) received the Ph.D.\ degree from Princeton University, Princeton, NJ. In 2004,
he joined the faculty of Northwestern University, Evanston, IL,
where he is currently a Professor in the Department of Electrical
Engineering and Computer Science. He has
been an Associate Editor of IEEE Transactions on Information Theory
and a Guest Editor of a Special Issue of IEEE Journal on Selected
Areas in Communications. He is an Editor of Foundations and Trends in
Communications and Information Theory.
Dr.~Guo received the IEEE Marconi Prize Paper Award in Wireless Communications in 2010 and a Best Paper Award at the 2017 IEEE Wireless Communications and Networking Conference.
He is also the recipient of the National Science Foundation Faculty Early
Career Development (CAREER) Award in 2007.
\end{IEEEbiography}

\begin{IEEEbiography}[{\includegraphics[width=1in,height=1.25in,clip,keepaspectratio]{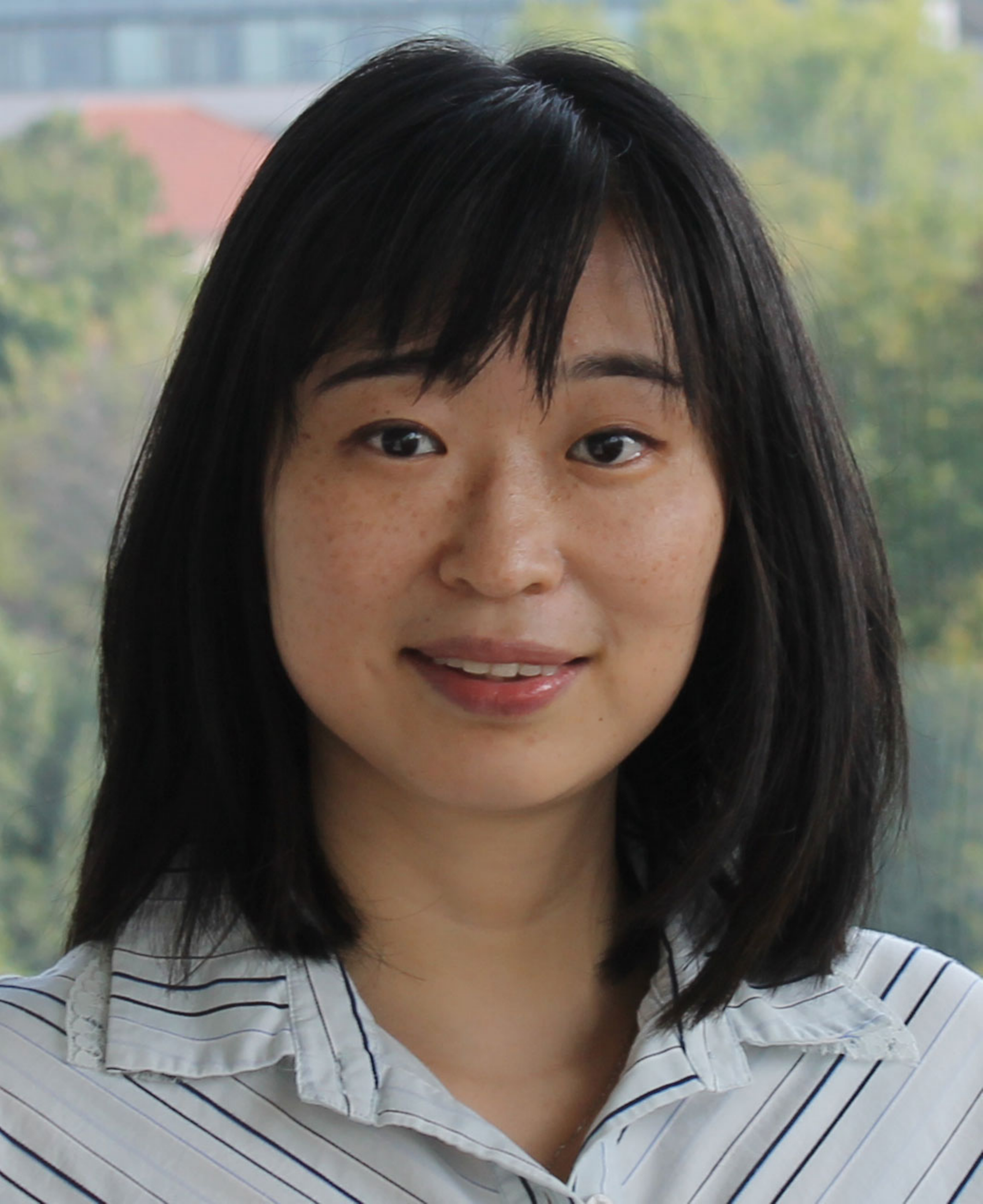}}]{Ermin Wei}
is currently an Assistant Professor at the EECS Dept of Northwestern University. She completed her PhD studies in Electrical Engineering and Computer Science at MIT in 2014, advised by Professor Asu Ozdaglar, where she also obtained her M.S.. She received her undergraduate triple degree in Computer Engineering, Finance and Mathematics with a minor in German, from University of Maryland, College Park. Wei has received many awards, including the Graduate Women of Excellence Award, second place prize in Ernst A. Guillemen Thesis Award and Alpha Lambda Delta National Academic Honor Society Betty Jo Budson Fellowship. Wei's research interests include distributed optimization methods, convex optimization and analysis, smart grid, communication systems and energy networks and market economic analysis.
\end{IEEEbiography}

\begin{IEEEbiography}[{\includegraphics[width=1in,height=1.25in,clip,keepaspectratio]{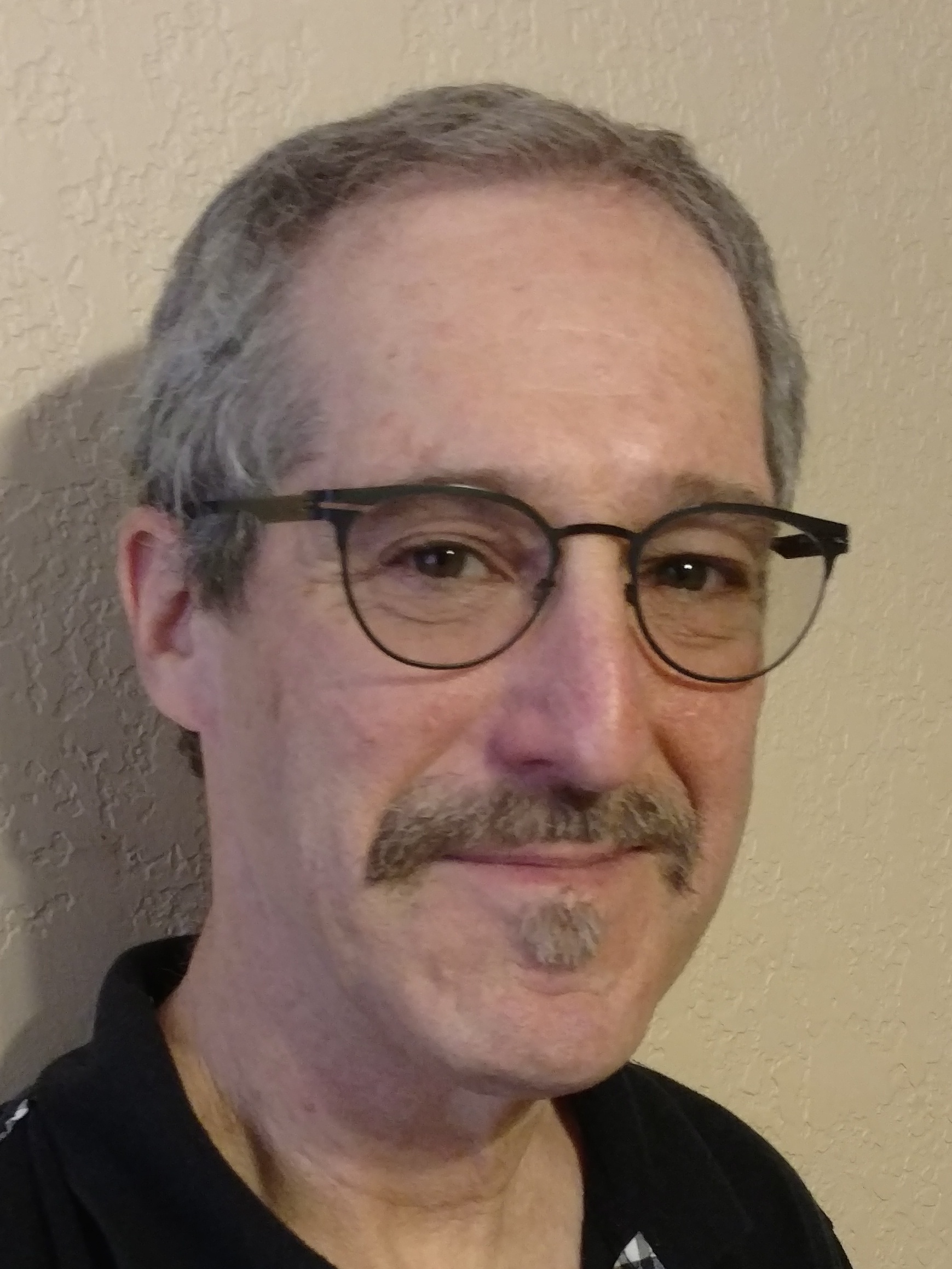}}]{Michael L. Honig}
(S'80-M'81-SM'92-F'97) received
the B.S. degree in electrical engineering from
Stanford University in 1977, and the M.S. and Ph.D.
degrees in electrical engineering from the University
of California, Berkeley, in 1978 and 1981, respectively.
He subsequently joined Bell Laboratories
in Holmdel, NJ, where he worked on local area
networks and voiceband data transmission. In 1983
he joined the Systems Principles Research Division
at Bellcore, where he worked on Digital Subscriber
Lines and wireless communications. Since the Fall
of 1994, he has been with Northwestern University where he is a Professor
in the Department of Electrical and Computer Engineering.
He has held several visiting scholar positions
and has also worked as a freelance trombonist.

Dr. Honig has served as an Editor for the IEEE Transactions on Information
Theory and the IEEE Transactions on Communications,
and as Guest Editor for several journals. He has also served
as a member of the Board of Governors for the Information
Theory Society. He is the recipient of a Humboldt Research
Award for Senior U.S. Scientists, and the co-recipient of the 2002 IEEE
Communications Society and Information Theory Society Joint Paper Award
and the 2010 IEEE Marconi Prize Paper Award.
\end{IEEEbiography}
\end{document}